\documentclass[final,3p,times,twocolumn]{elsarticle}

\usepackage{amssymb}
\usepackage{amsmath}
\usepackage{color}
\begin{document}

\begin{frontmatter}



\title{One-dimensional blast waves in a rarefied polyatomic gas with large bulk viscosity based on rational extended thermodynamics}


\author{Shigeru Taniguchi} %% Author name
\ead{taniguchi.shigeru@nihon-u.ac.jp}

\affiliation{organization={Department of Mathematical Information Engineering, Nihon University},%Department and Organization
            addressline={1-2-1 Izumi-cho}, 
            city={Narashino},
            postcode={275-8575}, 
            state={Chiba},
            country={Japan}}

%% Abstract
\begin{abstract}
The one-dimensional blast waves in a rarefied polyatomic gas with large bulk viscosity are studied based on rational extended thermodynamics (RET) with six independent fields: mass density, velocity, (equilibrium) pressure, and dynamic pressure.
First, by using the method of Lie group theory, we derive a similarity solution of blast waves induced by an intense point explosion. 
We discuss the deviation from the well-known Sedov-von Neumann-Taylor solution due to the dynamic pressure. 
Second, we analyze the time evolution by numerically solving the field equations of the RET theory directly for both cases of the intense explosion corresponding to the similarity solution and moderately strong explosion with generic temperature dependence of the bulk viscosity. 
It is shown that the prediction by the RET theory shows quite different behaviors from those by the system of the Euler equations, and also the Navier-Stokes-Fourier theory when the relaxation time for the dynamic pressure is considerable. 
\end{abstract}


% \begin{highlights}
% \item We quantitatively clarify the effect of the dynamic pressure on a plane blast wave induced by a point explosion in a polyatomic gas with large viscosity. 
% \item A similarity solution based on rational extended thermodynamics is derived, and the deviation from the Sedov-von Neumann-Taylor solution is discussed. 
% \item The difference between the predictions by rational extended thermodynamics and Navier-Stokes-Fourier theory becomes evident when the relaxation time is large enough. 
% \end{highlights}

%% Keywords
\begin{keyword}
Blast wave \sep Rational extended thermodynamics \sep Polyatomic gas \sep Similarity solution \sep Dynamic pressure \sep Bulk viscosity


\end{keyword}

\end{frontmatter}



\section{Introduction}

The blast wave has been attracting the attention of researchers because its highly non-equilibrium behaviors are physically interesting, and also important in many engineering applications~\cite{2010_Needham}. 
Classically, for an intense point explosion in a gas, the Sedov-von Neumann-Taylor (SvNT) self-similar solution~\cite{1993_Sedov} was derived based on the system of the Euler equations. 
Recently, the deviation from the SvNT solution for plane, cylindrical, and spherical blast waves has been studied in the case of monatomic gases~\cite{2021_Chakraborti,2021_Ganapa,2021_JoyPathakRajesh,2021_JoyRajesh,2022_KumarRajesh}. 
These studies emphasize the necessity of considering dissipative quantities, such as viscosity and heat conduction, through a comparison between the predictions by the SvNT solution, molecular dynamics simulations, and the Navier-Stokes-Fourier (NSF) theory. 

In the present study, we consider the blast waves in a polyatomic gas because it is known that the shock waves in a polyatomic gas show quite different features from those in a monatomic gas~\cite{1965_VincentiKruger}. 
The thickness of the steady plane shock wave can be much larger than the mean free path. 
Moreover, as the shock velocity increases from one, the profile of the shock wave structure changes from a nearly symmetric profile to an asymmetric profile and then to a profile composed of thin and thick layers. 
The Bethe-Teller theory~\cite{1941_BetheTeller} and Thermodynamics of Irreversible Processes~\cite{1963_deGrootMazur} (TIP) based on the local equilibrium assumption, including the NSF theory~\cite{1953_GilbargPaolucci}, can not explain these features in a unified way.  

Rational extended thermodynamics (RET)~\cite{1998_MullerRuggeri,2021_RuggeriSugiyama} is a phenomenological theory applicable beyond the local equilibrium assumption and is expected to describe highly non-equilibrium phenomena. 
In fact, the experimental data of the steady plane shock wave on the mass density profile in carbon dioxide gas are quantitatively explained for all three types of profiles~\cite{2014_PhysRevE} by the RET theory with 14 independent variables (RET$_{14}$)~\cite{2012_CMT}. 
It is also shown that the relaxation time for the dynamic pressure can be several orders of magnitude larger than other relaxation times. 

For polyatomic gases with the large relaxation time for the dynamic pressure, which corresponds to the large bulk viscosity, the predictions of the shock structure by RET$_{14}$ can be approximated to those by the simplified version of the RET theory~\cite{2014_PhysFluids,2016_IJNLM} (RET$_6$)~\cite{2012_PLA,2015_IJNLM,2016_Ruggeri,2018_BisiRuggeriSpiga}, which adopts only six independent fields: mass density $\rho$, velocity $\mathbf{v}$, (equilibrium pressure) $p$, and dynamic pressure $\Pi$.
The usefulness of the RET$_6$ theory is also confirmed by the fact that kinetic theory for a polyatomic gas reproduces the predictions by the RET$_6$ theory quantitatively even for high Mach numbers~\cite{2018_KosugeAoki,2019_KosugeKuoAoki}.

The next step is to analyze and clarify the effect of the dynamic pressure on the non-steady shock waves by using the RET$_6$ theory of polyatomic gases. 
For this purpose, the similarity solutions for the spherical and cylindrical shock waves have been derived and analyzed~\cite{2019_Nagaoka,2021_RdM}. 
The present study aims to analyze the one-dimensional (plane) blast waves based on the RET$_6$ theory in more detail. 

In the first part, by exploiting the method of group analysis~\cite{1995_DonatoOliveri,2000_DonatoRuggeri}, we analyze the one-dimensional blast wave induced by an intense explosion as done for the spherical and cylindrical shock waves~\cite{2019_Nagaoka,2021_RdM}. 
By deriving and numerically solving the similarity solution, we discuss the critical role of the dynamic pressure on the profiles and analyze the deviation from the classical SvNT solution~\cite{1993_Sedov} quantitatively. 

Furthermore, we will resolve the following two limitations for the similarity solution: 

(I) The similarity solution is valid only for extremely intense point explosion because the boundary conditions for the similarity solution are obtained from the Rankine-Hugoniot conditions in the strong shock limit. 

(II) The similarity solution assumes a special temperature dependence of the bulk viscosity. 
In the derivation of the similarity solution, the bulk viscosity $\nu$ is usually assumed to be independent of mass density $\rho$ and have the power-law type dependence on the temperature $T$~\cite{1953_GilbargPaolucci,2014_PhysFluids,2016_IJNLM,2012_Cramer}: 
\begin{equation}\label{mub}
\begin{split}
&\nu \propto T^{n}, 
\end{split}
\end{equation}
where the constant value of the exponent $n$ is chosen to fit the experimental data. 
For example, in the case of carbon dioxide gas, the value of $n$ is estimated as $n \approx -1.3$~\cite{2012_Cramer}. 
However, in the derivation of the similarity solution, the value of $n$ is uniquely determined from the requirements of the group analysis. 
In fact, the exponent of $n$ is determined as $n=1/6$ for the spherical case~\cite{2019_Nagaoka} and $n=0$ for the cylindrical case~\cite{2021_RdM}. 

In the second part of the present paper,  to resolve these limitations, we explicitly analyze the time evolution of the physical quantities by numerically solving the field equations of the RET$_6$ system directly for a generic strength of the explosion and a generic value of the exponent $n$. 
By comparing the prediction by the RET$_6$ theory with those by the NSF theory and the Euler equations, we clarify the typical feature of the prediction by the RET$_6$ theory and discuss the validity ranges of the conventional theories. 

\section{Basic equations}

In the present study, we focus on the one-dimensional motion along the $x$-axis where the velocity is expressed as $\mathbf{v} = (v, 0, 0)$ and all independent variables depend on the time $t$ and the position $x$ as $\rho(x,t)$, $v(x,t)$, $p(x,t)$, and $\Pi(x,t)$. 
The system of the field equations of the RET$_6$ theory is summarized as follows:~\cite{2012_PLA,2015_IJNLM,2016_Ruggeri,2018_BisiRuggeriSpiga}  
\begin{equation}\label{ET1D}
\begin{split}
&\frac{\partial \rho}{\partial t}+\frac{\partial}{\partial  x}( \rho v ) = 0,\\
&\frac{\partial \rho v}{\partial t} + \frac{\partial }{\partial x} (p + \Pi + \rho v^2 ) = 0,\\
&\frac{\partial}{\partial t} ( 2\rho \varepsilon +\rho v^2 ) + \frac{\partial}{\partial x}\left\{ 2 \rho \varepsilon v + 2(p+\Pi)v + \rho v^3 \right\} = 0,\\
&\frac{\partial}{\partial t}\left\{ 3(p+\Pi) +\rho v^2 \right\} + \frac{\partial}{\partial x}\left\{ 5(p+\Pi)v + \rho v^3 \right\}
= -\frac{3\Pi}{\tau},
\end{split}
\end{equation}
where $\varepsilon$ and $\tau$ are the specific internal energy and the relaxation time for the dynamic pressure. 
We adopt the caloric and thermal equations of state for a rarefied polyatomic gas: 
\begin{equation}\label{eqstate}
\begin{split}
&\varepsilon=\frac{1}{\gamma-1}\frac{p}{\rho}, \qquad p=\rho\frac{k_{B}}{m} T  
\end{split}
\end{equation}
with $\gamma$, $k_B$, and $m$ being, respectively, the ratio of the specific heats, the Boltzmann constant, and the mass of a molecule. 
In the present study, we also assume that the gas is polytropic, that is, the ratio of the specific heats $\gamma$ has a constant value independent of the temperature. 

By applying the so-called Maxwellian iteration~\cite{1956_IkenberryTruesdell}, the system of the RET$_6$ theory can be reduced to the system of the NSF theory only with the bulk viscosity, and the dynamic pressure $\Pi$ is expressed as  
\begin{equation}\label{eq:Maxwell_iteration}
    \Pi = - \left(\frac{5}{3}-\gamma \right) p \tau \frac{\partial v}{\partial x}. 
\end{equation}
In this sense, we understand that the RET$_6$ theory includes the NSF theory as a special case.  
From the comparison between the expression \eqref{eq:Maxwell_iteration} and the conventional NSF theory only with the bulk viscosity, we obtain the following relationship between the relaxation time for the dynamic pressure $\tau$ and the bulk viscosity $\nu$:  
\begin{equation}\label{mub_tau}
\begin{split}
&\nu=\left(\frac{5}{3}-\gamma \right) p \tau.
\end{split}
\end{equation}
Therefore, we can estimate the dependence of the relaxation time $\tau$ on the mass density $\rho$ and the pressure $p$ by inserting the power-law dependence \eqref{mub} into \eqref{mub_tau}. 

\section{Similarity solutions for an intense point explosion}
\label{sec:similarity_solution}

Let us consider an intense point explosion, such that a large amount of energy $E_0$ 
is released into a rarefied polyatomic gas instantaneously, and the perturbations propagate along the $x$-axis. 
It is assumed that the background of the gas is in equilibrium and at rest; $\rho=\rho_0$, $v=0$, $p=p_0$, and $\Pi=0$ with $\rho_0$ and $p_0$ being the mass density and pressure in the background state. 
From \eqref{mub}, \eqref{eqstate}, and \eqref{mub_tau}, we have the expression of the relaxation time $\tau$ in a generic state after the explosion as
\begin{equation}\label{tau0}
\tau(\rho, p)=\tau_0 \left(\frac{p}{p_{0}}\right)^{n-1} \left(\frac{\rho_0}{\rho}\right)^{n}, 
\end{equation}
where $\tau_0$ is the relaxation time evaluated in the background state: $\tau_0=\tau(\rho_0, p_0)$.

\subsection{System for similarity solutions}

As done in the spherical~\cite{2019_Nagaoka} and cylindrical~\cite{2021_RdM} cases, we derive the similarity solution for an intense point explosion for the present plane case. 
The energy $E_0$ within the position at the shock front $X (t)$ is given by~\cite{1993_Sedov}
\begin{equation}\label{E0_def}
\begin{split}
&E_0 = 2 \int_{0}^{X(t)} \left(\frac{1}{\gamma-1}p+\frac{1}{2}\rho v^{2} \right) dx,  
\end{split}
\end{equation}
where the blast wave induced by the explosion propagates in both positive and negative directions along the $x$-axis, but we focus on the one propagating in the positive $x$ direction for the similarity solution. 

Except for the functional form of the energy $E_0$, the calculations to derive the system of the similarity solution based on the Lie group theory~\cite{1995_DonatoOliveri,2000_DonatoRuggeri} are quite similar to the spherical~\cite{2019_Nagaoka} and cylindrical~\cite{2021_RdM} cases. 
We here only show the results of the derived similarity variables and the system for the similarity solution, and the details of the derivation are summarized in \ref{sec:derivation}. 

For obtaining the system of similarity solution in the plane case, the exponent $n$ is required to be
\begin{equation*}
n = -\frac{1}{2}, 
\end{equation*}
while $n=1/6$ is required in the spherical case ~\cite{2019_Nagaoka} and $n=0$ in the cylindrical case~\cite{2021_RdM}.  

We introduce the following dimensionless quantities for obtaining the similarity solution: 
\begin{equation}\label{eq:dimless_similarity}
\begin{split}
&\hat{\xi}=\left(\frac{\rho_{0}}{E_{0}}\right)^{\frac{1}{3}}\frac{x}{t^{\frac{2}{3}}}, \quad
\hat{\xi_0}=\left(\frac{\rho_{0}}{E_{0}}\right)^{\frac{1}{3}}\frac{X}{t^{\frac{2}{3}}}, \quad
\lambda=\frac{\hat{\xi}}{\hat{\xi_0}}, \\
&\rho = E_0 \, \hat{\xi}_{0}^3 \, \frac{t^2}{x^3} \hat{R}(\lambda), \quad
v=\frac{2}{3} \frac{x}{t} \left( \hat{V}(\lambda)+1 \right), \\
&p=\frac{4}{9} E_0 \, \hat{\xi}_{0}^3 \frac{1}{x} \hat{P}(\lambda), \quad
\Pi=\frac{4}{9} E_0 \, \hat{\xi}_{0}^3 \frac{1}{x} \hat{\Sigma}(\lambda).  
\end{split}
\end{equation} 
The system \eqref{ET1D} can be rewritten as follows: 
\begin{equation}\label{ET_result_2}
\begin{split}
&\hat{V} \frac{d \hat{R}}{d \ln \lambda} + \hat{R} \frac{d \hat{V}}{d \ln \lambda} 
= \hat{R}(2\hat{V}-1), \\
&\hat{V} \frac{d \hat{V}}{d \ln \lambda} + \frac{1}{\hat{R}}\frac{d}{d \ln \lambda}(\hat{P}+\hat{\Sigma}) \\
&= \frac{1}{R}(\hat{P}+\hat{\Sigma}) - \hat{V}\left(\hat{V} + \frac{1}{2}\right) + \frac{1}{2}, \\
&\hat{V} \frac{d \hat{P}}{d \ln \lambda} + \left\{ \gamma \hat{P} + (\gamma-1)\hat{\Sigma}\right\} \frac{d \hat{V}}{d \ln \lambda}\\
&= - (\gamma-1)(\hat{V}+1)\left(\hat{P} + \hat{\Sigma}\right), \\
&\hat{V} \frac{d \hat{\Sigma}}{d \ln \lambda} + \left\{ \left(\frac{5}{3}-\gamma\right)\hat{P}+\left(\frac{8}{3}-\gamma\right)\hat{\Sigma}\right\} \frac{d \hat{V}}{d \ln \lambda} \\
&= - \frac{\hat{\Sigma}}{\alpha}\sqrt{\frac{\hat{P}^3}{\hat{R}}} - (\hat{V}+1)\left( \frac{5}{3}-\gamma\right)(\hat{P}+\hat{\Sigma}), 
\end{split}
\end{equation}
where the dimensionless parameter $\alpha$ is given by
\begin{equation}\label{alpha}
\begin{split}
&\alpha = \frac{9}{4}\sqrt{\frac{p_0^3}{\rho_0}}\frac{\tau_0}{E_0 \hat{\xi}_0^3}. 
\end{split}
\end{equation}
It is noticeable that the similarity solution is characterized by only one dimensionless parameter $\alpha$, like in the spherical~\cite{2019_Nagaoka} and cylindrical~\cite{2021_RdM} cases.

\subsection{Boundary conditions}

Now we need to recall the theorem proved by Boillat and Ruggeri that a sub-shock (discontinuous part in the shock structure) appears when the shock velocity $s$ exceeds the maximum characteristic velocity in the unperturbed state~\cite{1998_BoillatRuggeri}. 
After the intense point explosion, we can expect that the velocity at the shock front is greater than the maximum characteristic velocity in the background state, and as a consequence, a sub-shock emerges. 
Therefore, the boundary condition for the ODE system \eqref{ET_result_2} can be obtained from the Rankine-Hugoniot conditions for the full system \eqref{ET1D} derived in~\cite{2014_PhysFluids}: 
\begin{equation*}
\begin{split}
&\frac{\rho_{*}}{\rho_{0}}=\frac{4 \gamma M^{2}_{0}}{\gamma M^{2}_{0}+5},\\
&\frac{v_{*}}{c_{0}}=\frac{3 \gamma M^{2}_{0}-5}{4 \gamma M_{0}}, \\
&\frac{p_{*}}{p_{0}}=\frac{9 \gamma^{2}(\gamma-1)M^{4}_{0}+2 \gamma(19-3 \gamma)M^{2}_{0}-15(\gamma-1)}{40+8 \gamma M^{2}_{0}}, \\
&\frac{\Pi_{*}}{p_{0}}=(5-3 \gamma)\frac{3 \gamma^{2}M^{4}_{0}- 2 \gamma M^2_{0}-5}{40+8 \gamma M^{2}_{0}}, 
\end{split}
\end{equation*}
where the quantities with subscript $*$ represent the ones evaluated in the state just after the sub-shock, and $M_0$ represents the Mach number in the unperturbed state defined by
\begin{equation*}
M_0 = \frac{s-v_0}{c_0} 
\end{equation*} 
with $v_0$ and $c_0$ being the velocity and sound velocity in the unperturbed state. 
In the present case, $c_0=\sqrt{\gamma p_0/\rho_0}$ and $v_0=0$ holds. 
In the strong shock limit, the Rankine-Hugoniot conditions become
\begin{equation} \label{ET_RH_s}
\begin{split}
&\rho_{*}=4 \rho_{0}, \quad
v_{*}=\frac{3}{4}s, \quad \\
&p_*=\frac{9(\gamma-1)}{8}s^{2}\rho_{0}, \quad
\Pi_{*}=\frac{3(5-3 \gamma)}{8}s^{2}\rho_{0}.
\end{split}
\end{equation}
From \eqref{eq:dimless_similarity}, the position of the shock front $X(t)$ and the velocity of the shock front ${s}$ are obtained as
\begin{equation}\label{Rs}
\begin{split}
X(t) = \hat{\xi}_{0} \left(\frac{E_{0}}{\rho_{0}} \right)^{\frac{1}{3}} t^{\frac{2}{3}}, \quad
s=\frac{2}{3} \hat{\xi}_{0} \left(\frac{E_{0}}{\rho_{0}} \right)^{\frac{1}{3}} t^{-\frac{1}{3}}. 
\end{split}
\end{equation}
From \eqref{eq:dimless_similarity}, \eqref{ET_RH_s} and \eqref{Rs}, we have the boundary conditions at the shock front as 
\begin{equation}\label{eq:boundary}
\begin{split}
&\hat{R}(1)=4, \quad \hat{V}(1)=-\frac{1}{4}, \quad \\
&\hat{P}(1)=\frac{9(\gamma-1)}{8}, 
\quad \hat{\Sigma}(1)=\frac{3(5-3 \gamma)}{8}. 
\end{split}
\end{equation}

\subsection{Behavior predicted by the similarity solution}
\label{sec:spherical_shock}

The derived system \eqref{ET_result_2} under the boundary conditions \eqref{eq:boundary} can be solved numerically by using usual solvers for ODEs. 
Recalling the relation $\lambda=x/X$, we can relate the dimensionless quantities with the similarity solutions as follows: 
\begin{equation} \label{ET_RH_tilde}
\begin{split}
&\tilde{\rho}\equiv\frac{\rho}{\rho_*}=\frac{1}{\lambda^3}\frac{\hat{R}(\lambda)}{\hat{R}(1)}, \quad
\tilde{v}\equiv\frac{v}{v_*}=\lambda\frac{\hat{V}(\lambda)+1}{\hat{V}(1)+1}, \\
&\tilde{p}\equiv\frac{p}{p_*}=\frac{1}{\lambda}\frac{\hat{P}(\lambda)}{\hat{P}(1)}, \quad 
\tilde{\Pi}\equiv\frac{\Pi}{p_*}=\frac{1}{\lambda}\frac{\hat{\Sigma}(\lambda)}{\hat{P}(1)}. 
\end{split}
\end{equation}

\begin{figure*}[h!]%% placement specifier
\centering%% For centre alignment of image.
\includegraphics[width=0.33\linewidth]{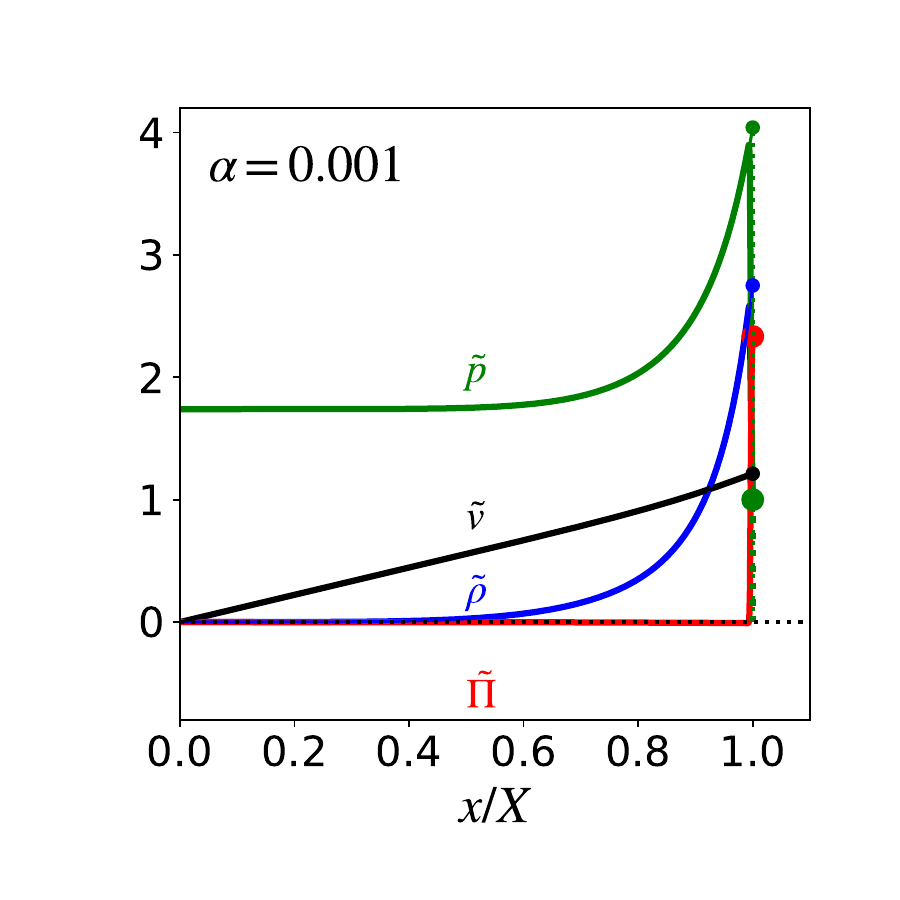}
\includegraphics[width=0.33\linewidth]{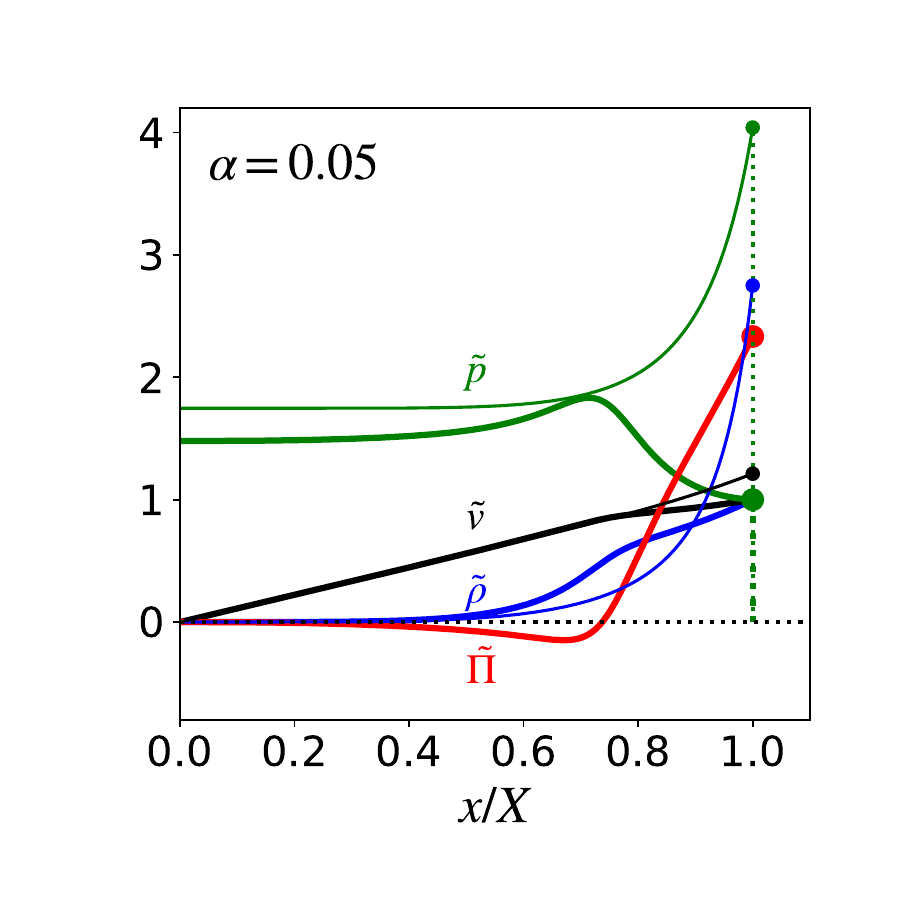}
\includegraphics[width=0.33\linewidth]{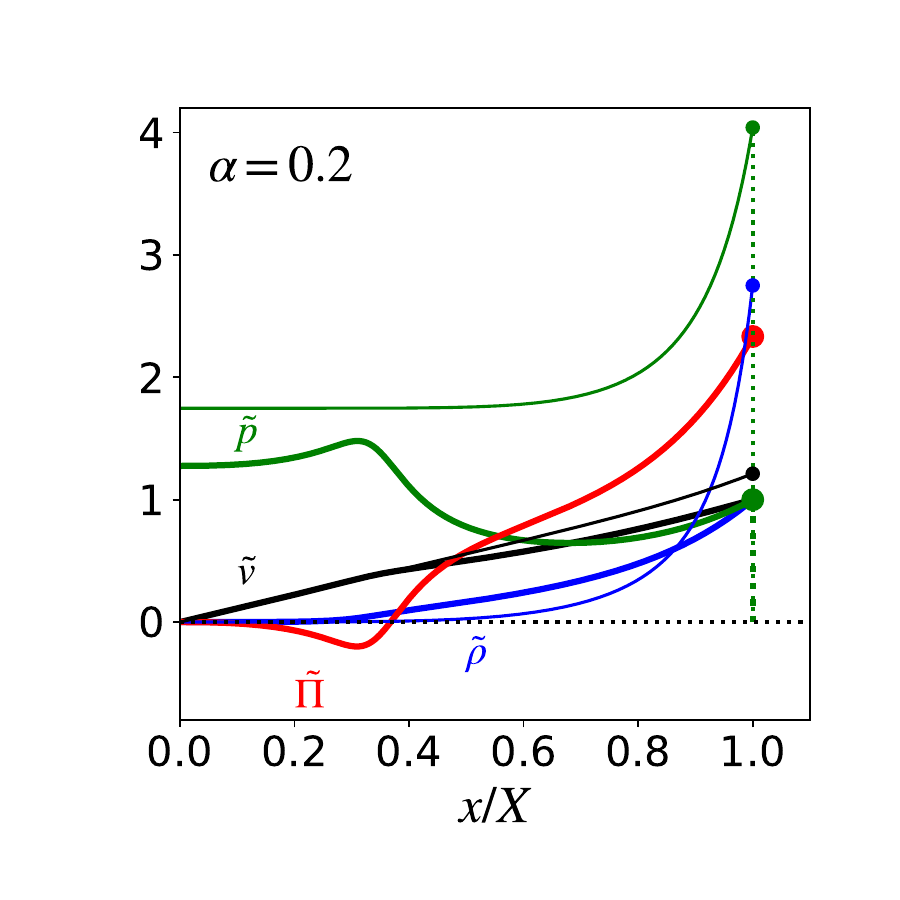}
\caption{Profiles of the physical quantities predicted by the similarity solutions by RET$_6$ (Think curves) and the SvNT solution (thin curves) for $\gamma=1.2$ for $\alpha=0.001$ (left), $\alpha=0.05$ (center), and $\alpha=0.2$ (right). 
The dots indicate the boundary values at the shock front, and in the case of the RET$_6$ theory, from the definition \eqref{ET_RH_tilde}, $\tilde{\rho}=\tilde{v}=\tilde{p}=1$ at the boundary. }
\label{fig:similarity_gamma1_2}
\end{figure*}

Figure \ref{fig:similarity_gamma1_2} shows the profile of the physical quantities predicted by the RET$_6$ theory in the case of $\gamma=1.2$ and for several values of the dimensionless parameter $\alpha$, namely, $\alpha=$ $0.001$, $0.05$, and $0.2$. 
The corresponding profiles predicted by the SvNT solution~\cite{1993_Sedov} are also depicted in Figure \ref{fig:similarity_gamma1_2} for comparison. 
We see that if the value of $\alpha$ is small enough, the prediction by the RET$_6$ theory agrees with that by the SvNT solution.  
It is noticeable that the RET$_6$ theory can explain a steep but continuous change at the peak of the physical quantities near the shock front at $x/X=1$, while the SvNT solution describes it as a discontinuous jump. 

From Figure \ref{fig:similarity_gamma1_2}, we also confirm that if the value of $\alpha$ becomes larger, the strength of the peak decreases, and the difference between the similarity solution based on the RET$_6$ theory and the SNT solution becomes more evident. 
As we increase the value of $\alpha$ more, we have completely different behavior from the SvNT solution, that is, the peaks of the (equilibrium) pressure and dynamic pressure appear far from the shock front. 

\section{Time evolution for a point explosion}

In this section, we numerically solve the system of the field equations \eqref{ET1D} directly. 
For the initial conditions, corresponding to the point explosion at $x=0$,  we assume the uniform mass density $\rho_0$, zero velocity $v_0=0$, zero dynamic pressure $\Pi = 0$, and the Gaussian-type initial pressure:  
\begin{equation}
    p (x, 0) = p_0 + (p_I - p_0) \exp\left(-\frac{x^2}{L^2}\right), 
\end{equation}
where $p_I$ is the pressure at the peak ($x=0$) and $L$ characterizes the thickness of the initial pressure profile. 
The relaxation time for the dynamic pressure can also be rewritten as
\begin{equation}\label{eq:tauL}
\tau(\rho, p)=\tau_I \left(\frac{p}{p_{I}}\right)^{n-1} \left(\frac{\rho_0}{\rho}\right)^{n}, 
\end{equation}
where $\tau_I$ is the relaxation time at the peak ($x=0$): $\tau_I=\tau(\rho_0, p_I)$. 

For convenience, we introduce the following dimensionless quantities: 
\begin{equation}\label{eq:dimensionless}
    \begin{split}
        &\hat{\rho} \equiv \frac{\rho}{\rho_0}, \quad
        \hat{v} \equiv \frac{v}{c_I}, \quad 
        \hat{p} \equiv \frac{p}{p_I}, \quad
        \hat{\Pi} \equiv \frac{\Pi}{p_I}, \\
        &\hat{t} \equiv \frac{t}{t_c}, \quad 
        \hat{x} \equiv \frac{x}{c_I t_c}, \quad 
        \hat{\tau}_I \equiv \frac{\tau_I}{t_c}, \quad
         \hat{L} \equiv \frac{L}{c_I t_c}, 
    \end{split}
\end{equation}
where $t_c$ is an arbitrary characteristic time for numerical computations and $c_I$ is the sound velocity evaluated at the peak of the initial pressure $c_I = \sqrt{\gamma p_I/\rho_0}$. 
The initial conditions can be summarized as follows: 
\begin{equation}
    \begin{split}
        &\hat{\rho} (\hat{x}, 0) = 1, \quad
        \hat{v} (\hat{x}, 0) = 0, \\
        &\hat{p} (\hat{x}, 0) = \hat{p}_0 + (1-\hat{p}_0)\exp\left(-\frac{\hat{x}^2}{\hat{L}^2} \right), \quad
        \hat{\Pi}(\hat{x}, 0) = 0, 
    \end{split}
\end{equation}
and we adopt $\hat{L}=0.2$ throughout this paper. 

The system of field equations \eqref{ET1D} is rewritten in terms of the dimensionless quantities as 
\begin{equation}\label{ET1D_dimless}
\begin{split}
&\frac{\partial \hat{\rho}}{\partial \hat{t}}+\frac{\partial}{\partial  \hat{x}}( \hat{\rho} \hat{v} ) = 0,\\
&\frac{\partial \hat{\rho} \hat{v}}{\partial \hat{t}} + \frac{\partial }{\partial \hat{x}} \left(\frac{\hat{p} + \hat{\Pi}}{\gamma} + \hat{\rho} \hat{v}^2 \right) = 0,\\
&\frac{\partial}{\partial \hat{t}} \left( \frac{2 \hat{p}}{\gamma(\gamma-1)} + \hat{\rho} \hat{v}^2 \right) + \frac{\partial}{\partial \hat{x}}\left\{ 2 \left( \frac{\hat{p}}{\gamma-1} + \frac{\hat{\Pi}}{\gamma} \right)\hat{v} + \hat{\rho} \hat{v}^3 \right\} = 0,\\
&\frac{\partial}{\partial \hat{t}}\left\{ \frac{3(\hat{p}+\hat{\Pi})}{\gamma} +\hat{\rho} \hat{v}^2 \right\} + \frac{\partial}{\partial \hat{x}}\left\{ \frac{5\left(\hat{p}+\hat{\Pi}\right)\hat{v}}{\gamma} + \hat{\rho} \hat{v}^3 \right\}\\
&\quad = -\frac{3 \hat{\rho}^n \hat{p}^{1-n} \hat{\Pi}}{\gamma \hat{\tau}_I}.
\end{split}
\end{equation}
Similarly, the equation of the NSF theory with only bulk viscosity \eqref{eq:Maxwell_iteration} becomes
\begin{equation}\label{NSF1D_dimless}
\hat{\Pi} = - \left(\frac{5}{3}-\gamma \right) \hat{\tau}_I \frac{\hat{p}^n}{\hat{\rho}^n} \frac{\partial \hat{v}}{\partial \hat{x}}. 
\end{equation}

For solving the field equations of the RET$_6$ theory~\eqref{ET1D_dimless}, we employ the Uniformly accurate Central Scheme of order 2 (UCS2)~\cite{2000_LiottaRomanoRusso}, which is proposed to solve the hyperbolic balance laws with stiff production terms and its usefulness has been tested in the analysis of the shock structure problem with sub-shocks in several hyperbolic systems~\cite{2006_MentrelliRuggeri,2017_IJNLM,2017_ConfortoMentrelliRuggeri,2022_PoF,2024_CAMC}. 

The system of the Euler equations can also be solved by UCS2 by dropping the term related to the dimensionless dynamic pressure $\hat{\Pi}$ in \eqref{ET1D_dimless}. 
For solving the field equations of the NSF theory \eqref{ET1D_dimless}$_{1-3}$ and \eqref{NSF1D_dimless}, we adopt the implicit Euler method~\cite{2020_FerzigerPericStreet}.

\subsection{Case 1. $n = -1/2$ and $\hat{p}_0 = 10^{-4}$}

\begin{figure*}[h!]
\begin{center}
    \includegraphics[width=0.33\linewidth]{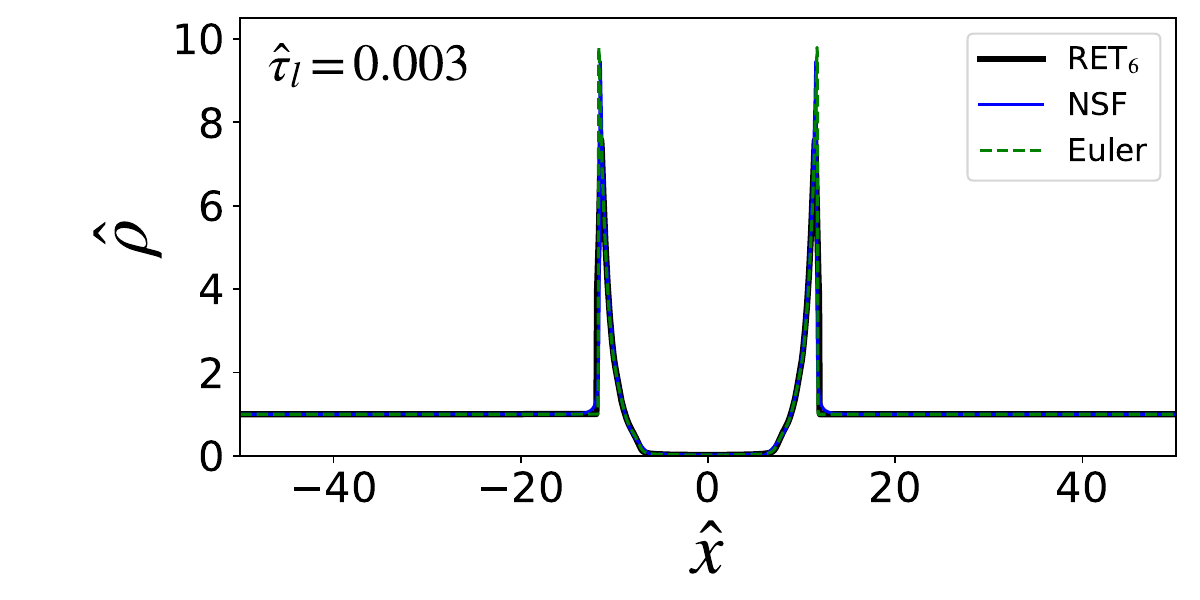}    
    \includegraphics[width=0.33\linewidth]{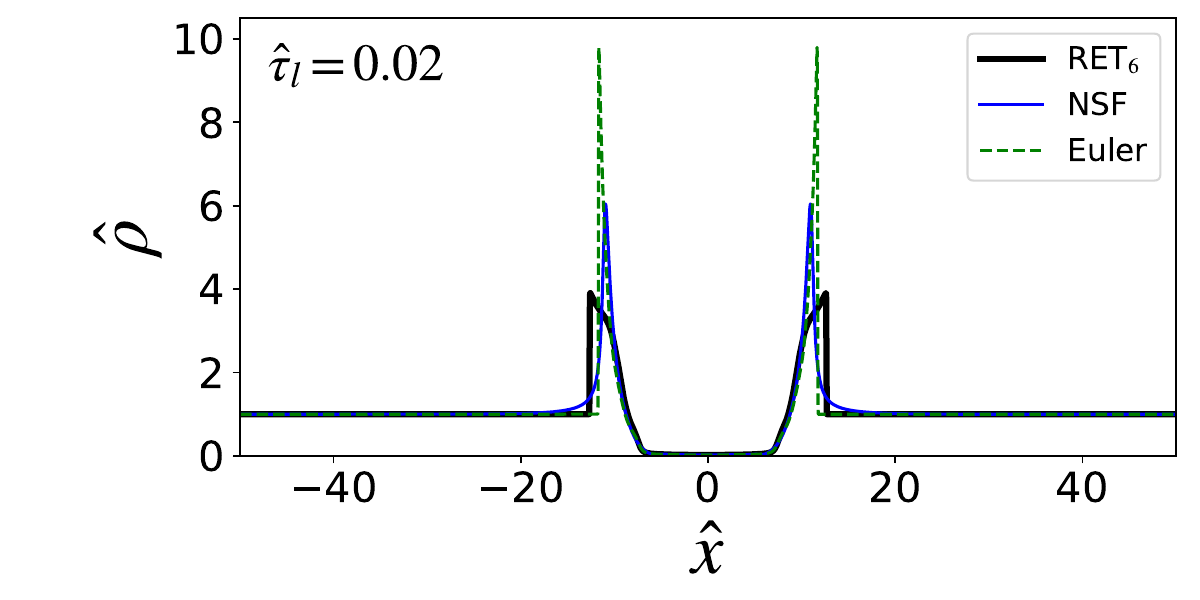}    
    \includegraphics[width=0.33\linewidth]{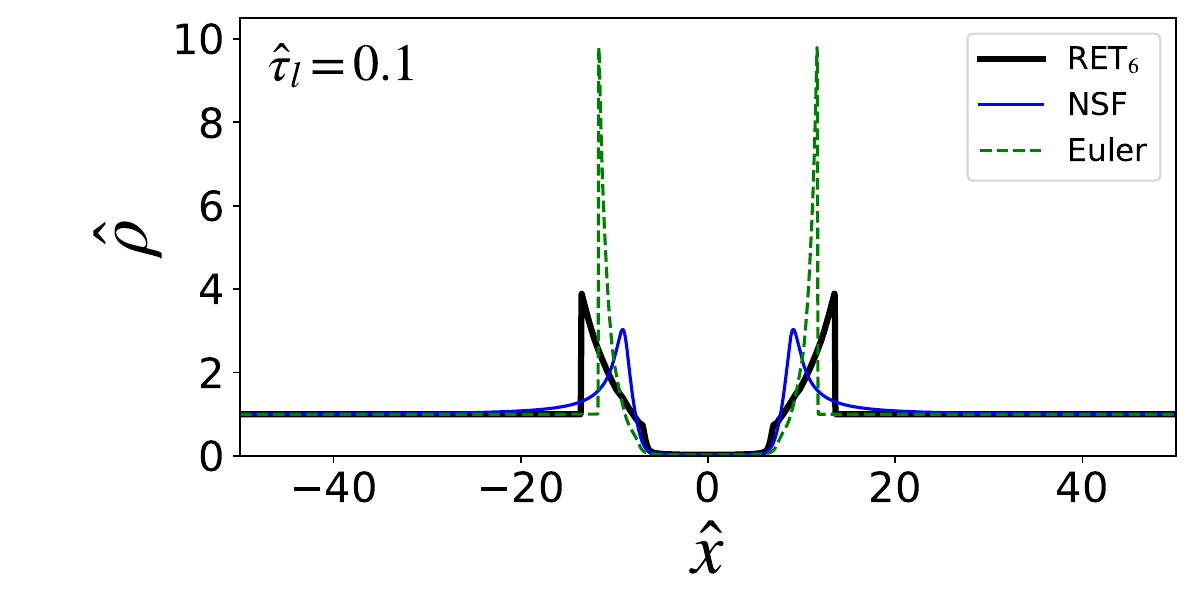}    
    \includegraphics[width=0.33\linewidth]{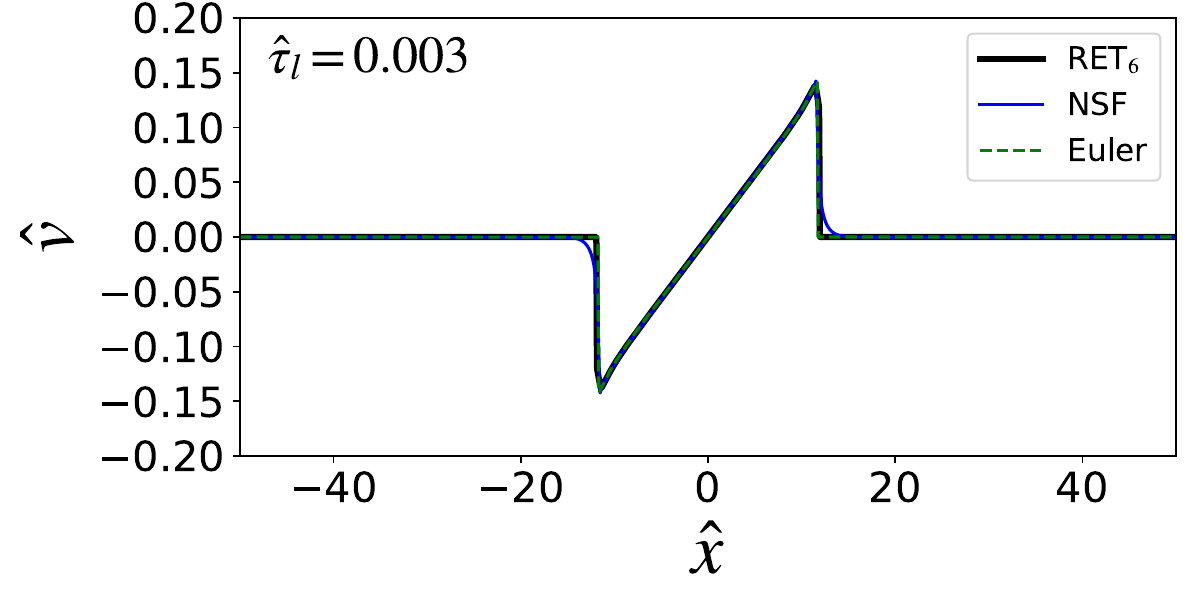}    
    \includegraphics[width=0.33\linewidth]{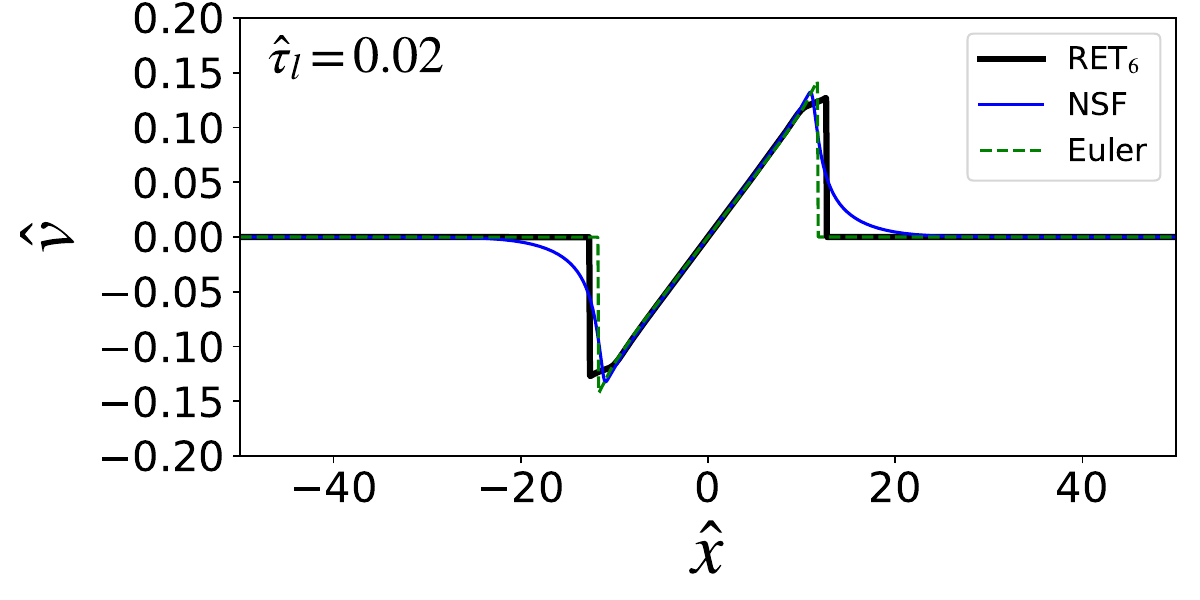}    
    \includegraphics[width=0.33\linewidth]{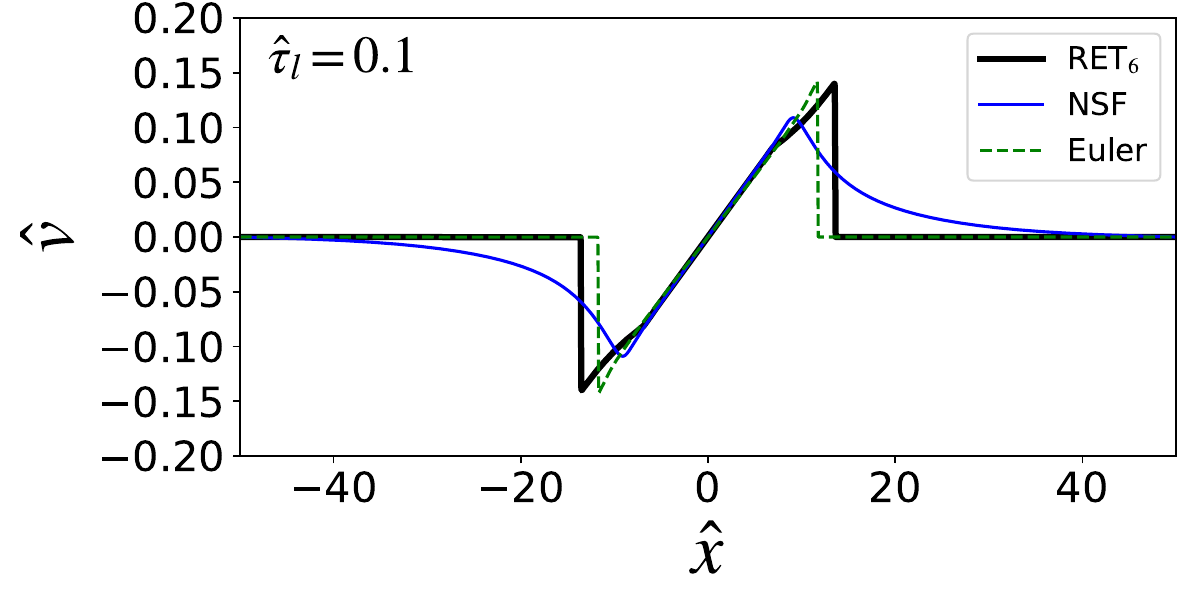}    
    \includegraphics[width=0.33\linewidth]{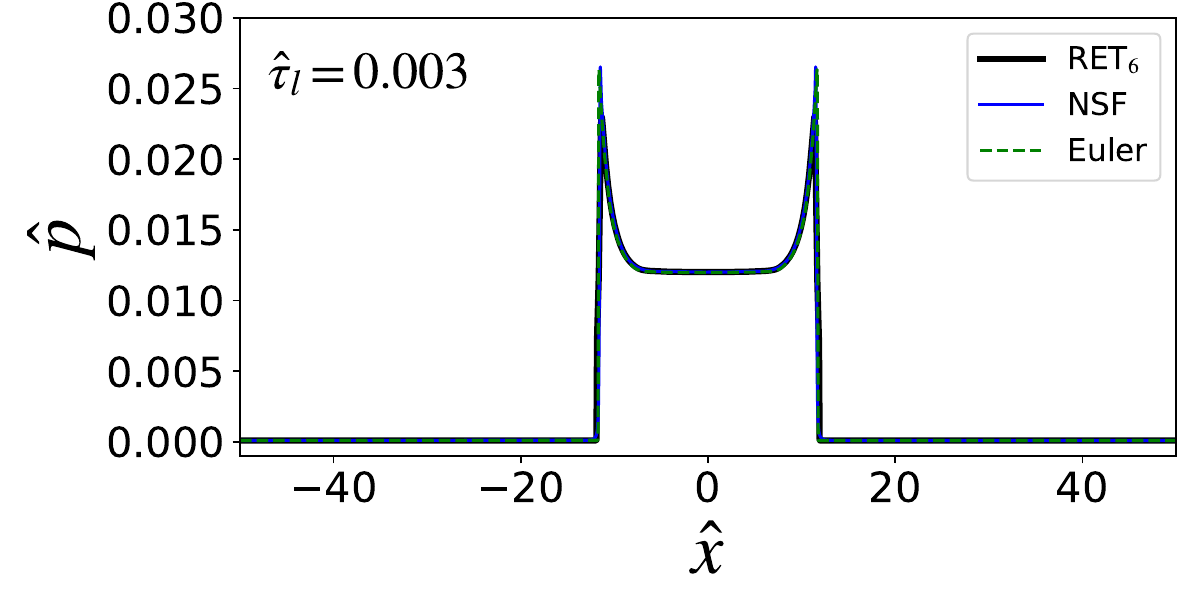}    
    \includegraphics[width=0.33\linewidth]{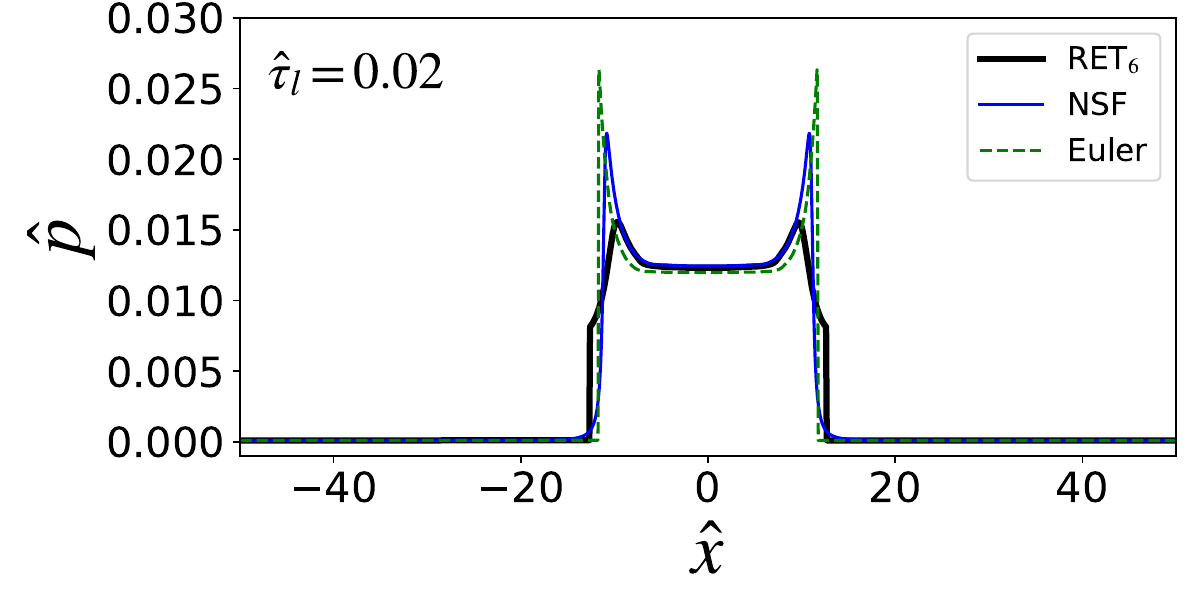}    
    \includegraphics[width=0.33\linewidth]{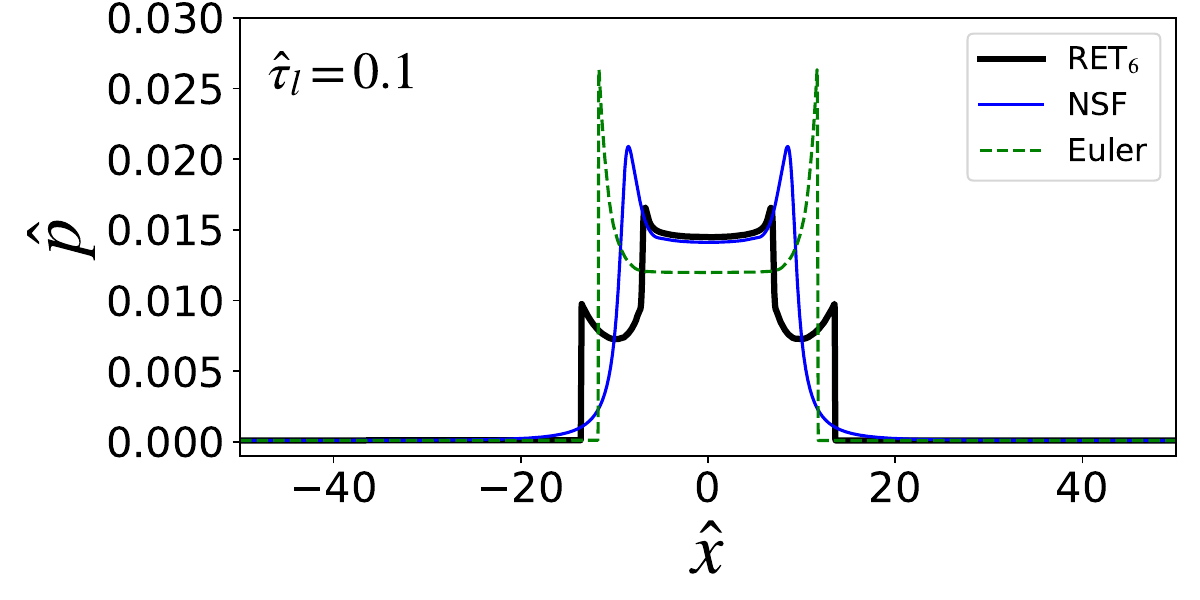}    
    \includegraphics[width=0.33\linewidth]{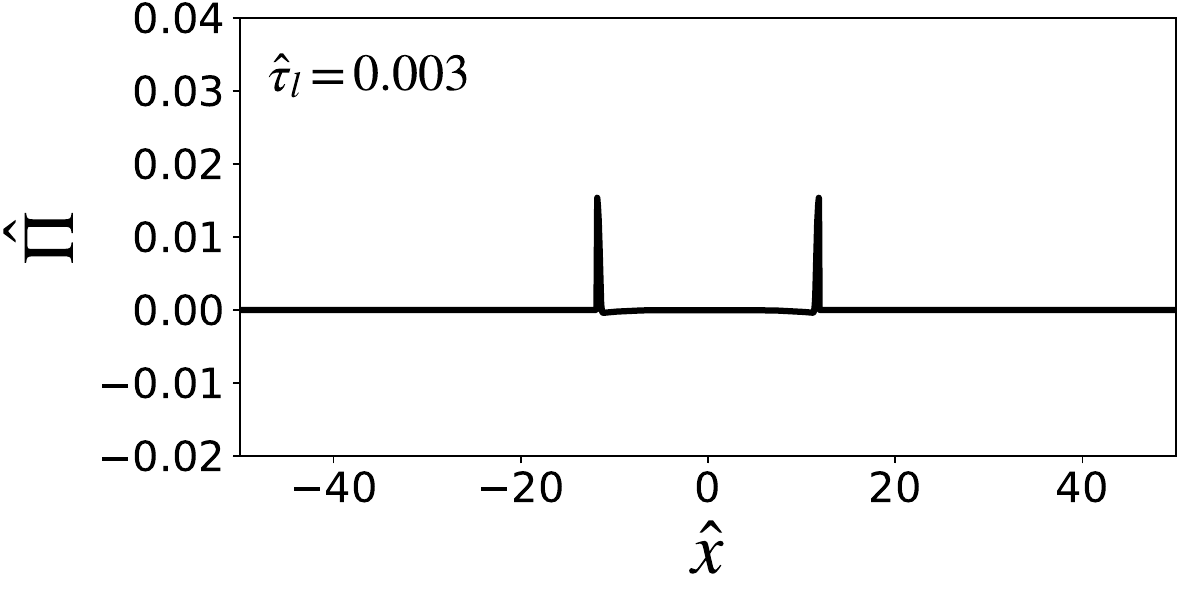}    
    \includegraphics[width=0.33\linewidth]{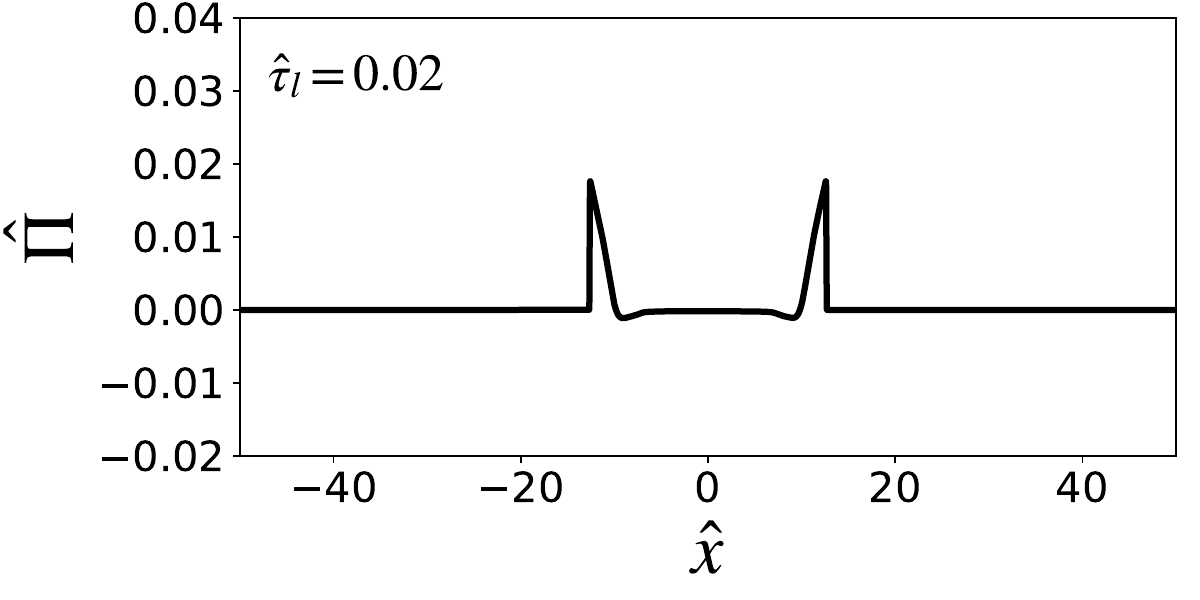}    
    \includegraphics[width=0.33\linewidth]{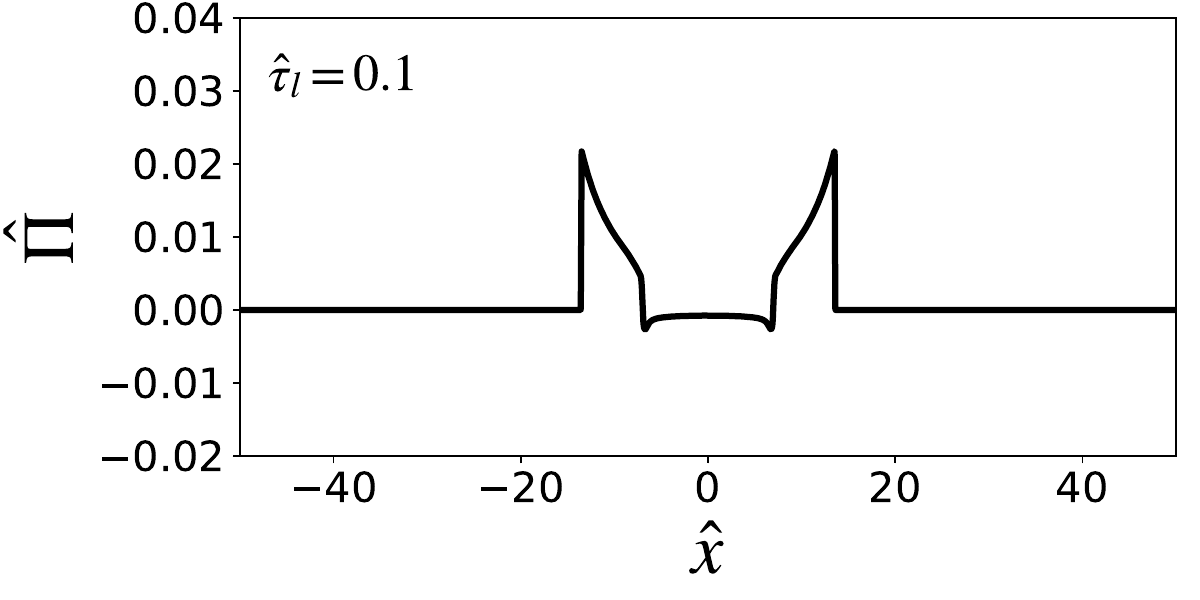}    
    \caption{Profiles of the physical quantities at $\hat{t} = 50$ predicted by the RET$_6$ theory (black thick curves), the NSF theory (blue thin curves), and the system of Euler equations (green dashed curves). 
    The parameters are $\gamma=1.2$, $\hat{L}=0.2$, $\hat{p}_0 = 10^{-4}$, and $n=-1/2$.
    The dimensionless relaxation times are $\hat{\tau}_l=0.003$ (left column), $\hat{\tau}_l=0.02$ (center column), and $\hat{\tau}_l=0.1$ (right column). }
    \label{fig:gamma1_2_pratio10000_th50}
\end{center}
\end{figure*}

First, we check the validity of the similarity solution by taking a large initial pressure ratio $\hat{p}_0=10^{-4}$ with $n=-1/2$. 
We show the profiles of the dimensionless physical quantities at $\hat{t}=50$ for several dimensional relaxation times $\hat{\tau}_I =$ $0.003$, $0.02$, and $0.1$ in Figure \ref{fig:gamma1_2_pratio10000_th50}. 
For comparison, the predictions by the NSF theory and also the system of the Euler equations are also shown in Figure \ref{fig:gamma1_2_pratio10000_th50}.

Although we can not take exactly the strong shock limit in the present numerical calculations, we confirm that the limiting value of the mass density $\hat{\rho}*$ becomes near $4$ as predicted by \eqref{ET_RH_s} for all cases. 
Also, the peak of the mass density in the system of the Euler equations is almost $10$ times larger than the unperturbed mass density predicted by the usual Rankine-Hugoniot conditions in the strong shock limit~\cite{1993_Sedov} with $\gamma = 1.2$. 
Therefore, we consider that the present situation can be compared with the similarity solution derived in Section~\ref{sec:similarity_solution}. 

Inserting \eqref{tau0}, \eqref{Rs}$_1$, \eqref{eq:tauL} and \eqref{eq:dimensionless} into \eqref{alpha}, we can rewrite the dimensionless parameter $\alpha$ in terms of the dimensionless quantities as
\begin{equation}\label{alpha2}
\alpha = \frac{9}{4} \gamma^{-\frac{3}{2}} \hat{\tau}_I \frac{\hat{t}^2}{\hat{X}^3}, 
\end{equation}
where $\hat{X}$ is the dimensionless position of the shock front $\hat{X} = X/(c_I t_c)$. 
From Figure \ref{fig:gamma1_2_pratio10000_th50}, we read the value of $\hat{X}$ at $\hat{t}=50$ and by inserting these values into \eqref{alpha2}, we estimate the value of dimensionless parameter $\alpha$ as follows: 
\begin{itemize}
    \item $\hat{X} = 11.8$ and $\alpha=0.0078$ in the case of $\hat{\tau}_I=0.003$. 
    \item $\hat{X} = 12.6$ and $\alpha=0.043$ in the case of $\hat{\tau}_I=0.02$. 
    \item $\hat{X} = 13.7$ and $\alpha=0.17$ in the case of $\hat{\tau}_I=0.1$.
\end{itemize}
We confirm the validity of the results of the similarity solution depending on the dimensionless parameter $\alpha$. 
If the value of dimensionless relaxation time $\hat{\tau}_I$ is small enough, which corresponds to a small value of $\alpha$, the predictions by these theories agree with each other. 
If we increase the value of $\hat{\tau}_I$, which implies the increase of $\alpha$, the RET$_6$ theory predicts qualitatively different behavior from the SvNT solution. 

Remarkably, the predictions by the RET$_6$ and NSF theories are also quite different from each other if $\hat{\tau}_I$ becomes large ($\alpha$ becomes large). 
Because the NSF theory is parabolic, the predicted profile is always smooth, and no sub-shock appears. 
Moreover, the NSF theory predicts much thicker profiles near the shock front than the RET$_6$ theory. 

\subsection{Case 2. $n = -1.3$ and $\hat{p}_0 = 10^{-2}$}

\begin{figure*}[h!]
\begin{center}
    \includegraphics[width=0.33\linewidth]{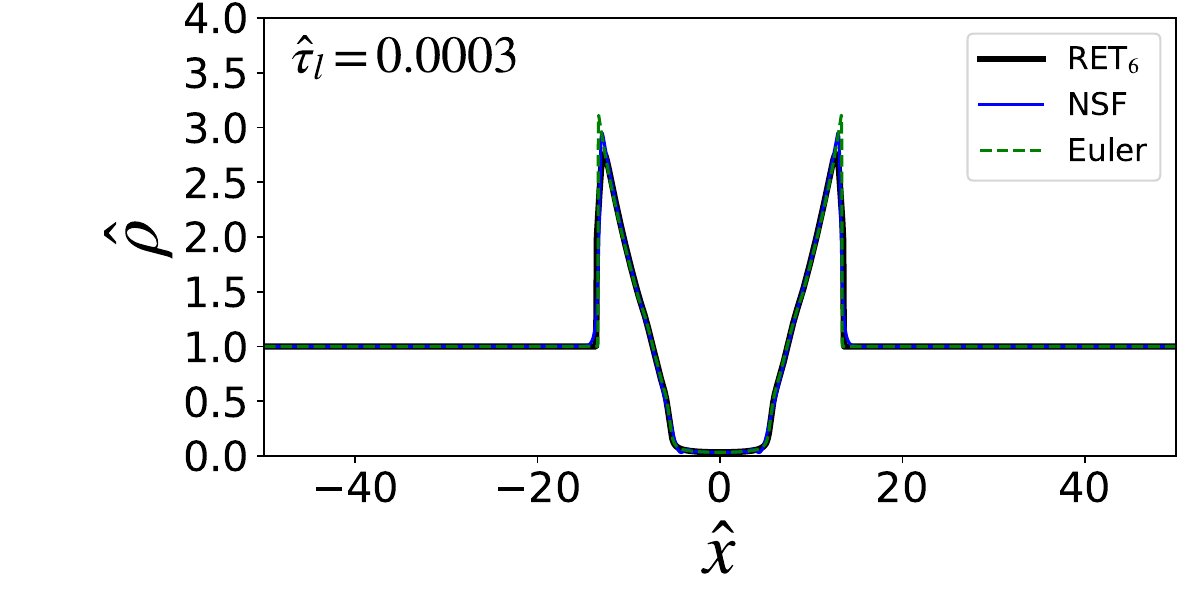}    
    \includegraphics[width=0.33\linewidth]{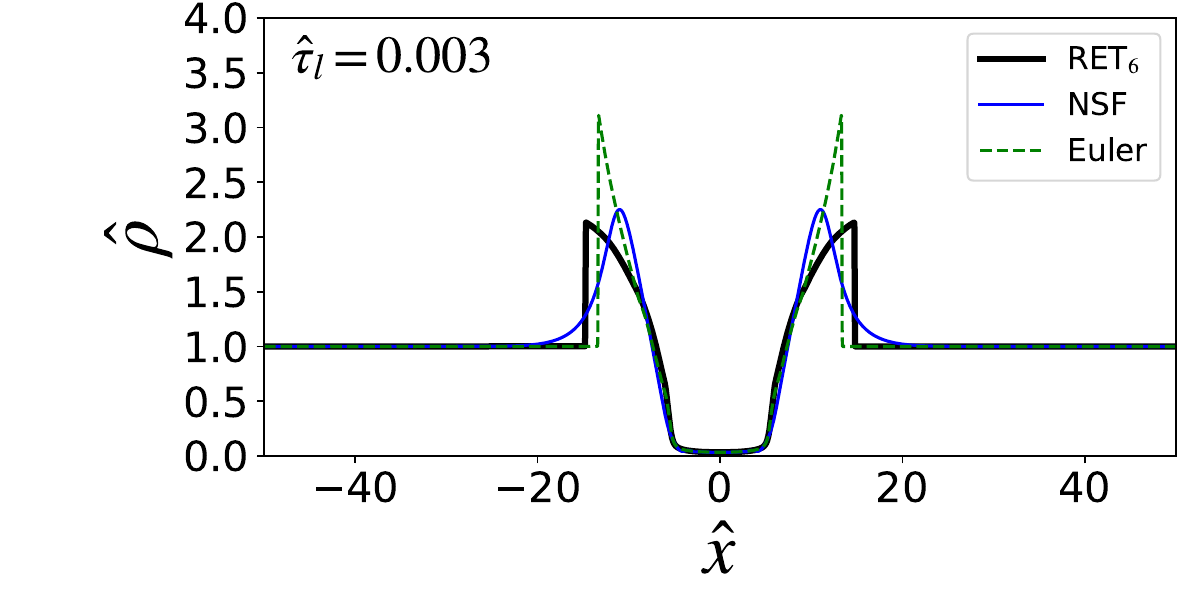}    
    \includegraphics[width=0.33\linewidth]{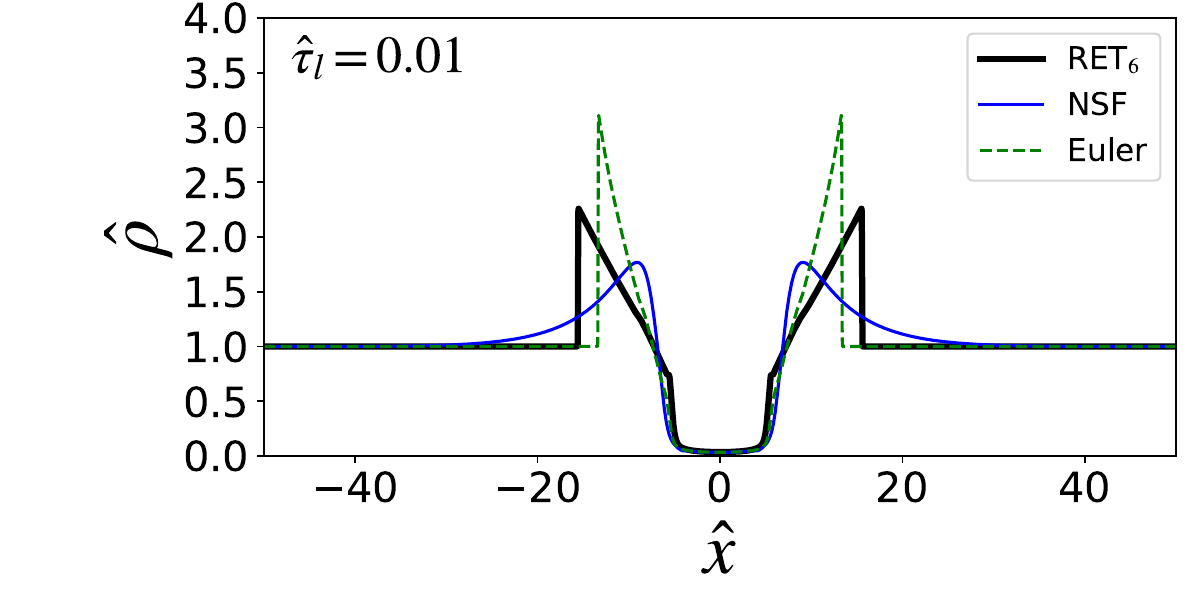}    
    \includegraphics[width=0.33\linewidth]{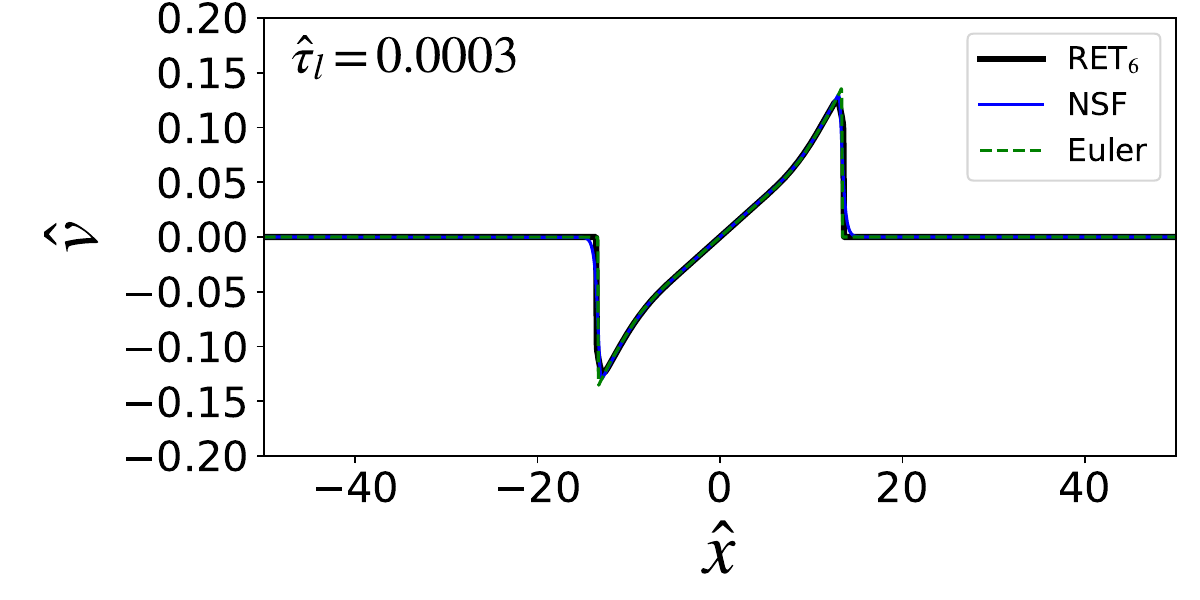}    
    \includegraphics[width=0.33\linewidth]{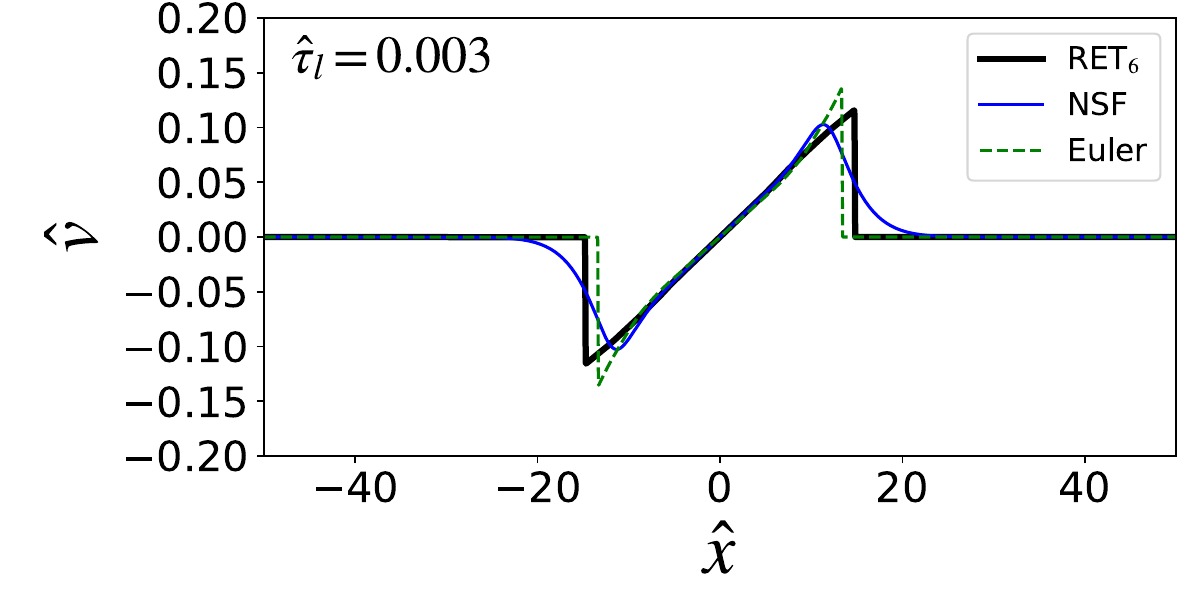}    
    \includegraphics[width=0.33\linewidth]{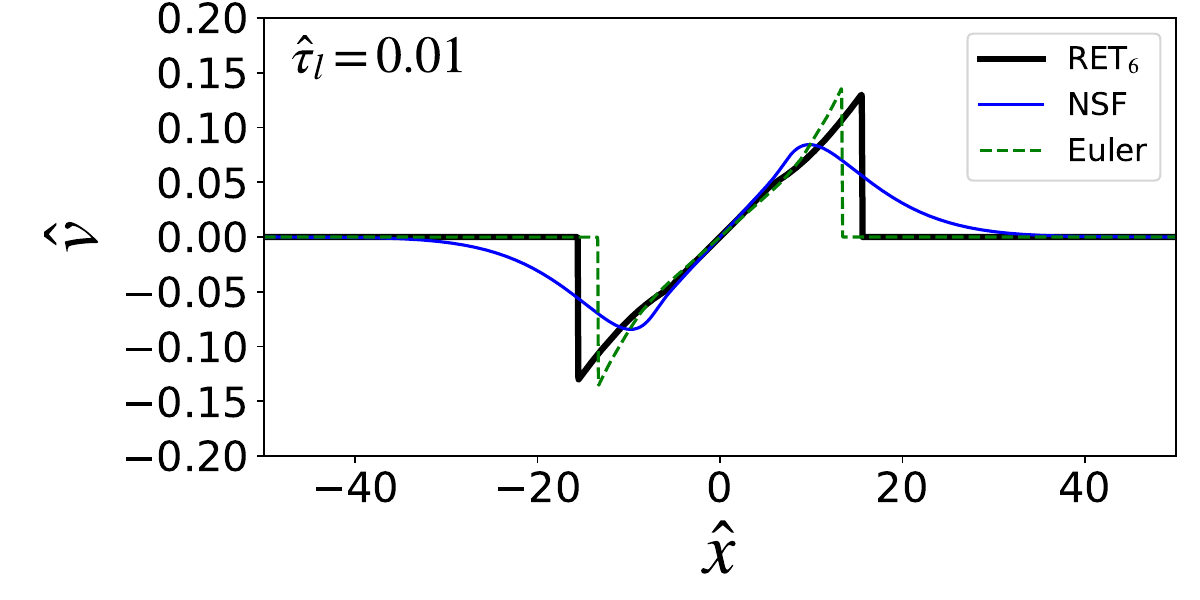}    
    \includegraphics[width=0.33\linewidth]{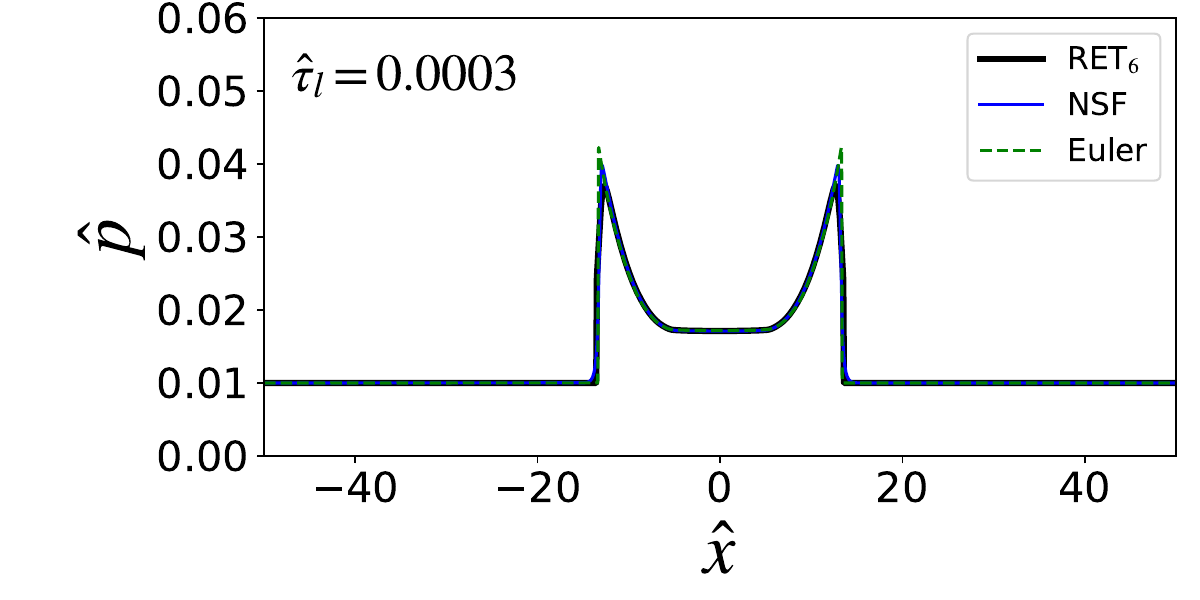}    
    \includegraphics[width=0.33\linewidth]{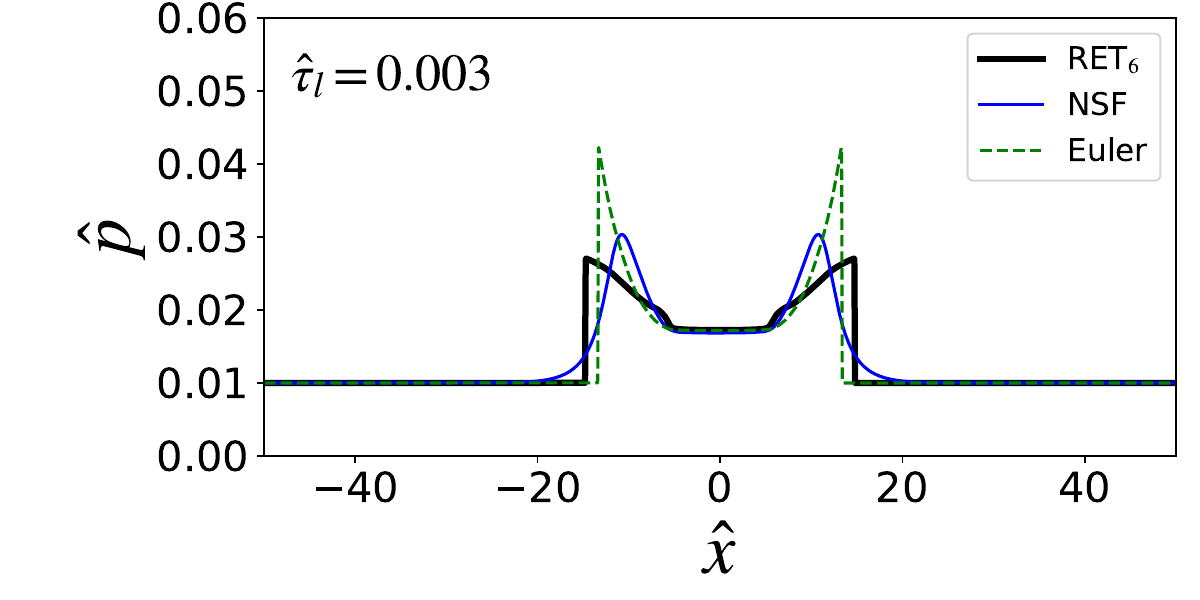}    
    \includegraphics[width=0.33\linewidth]{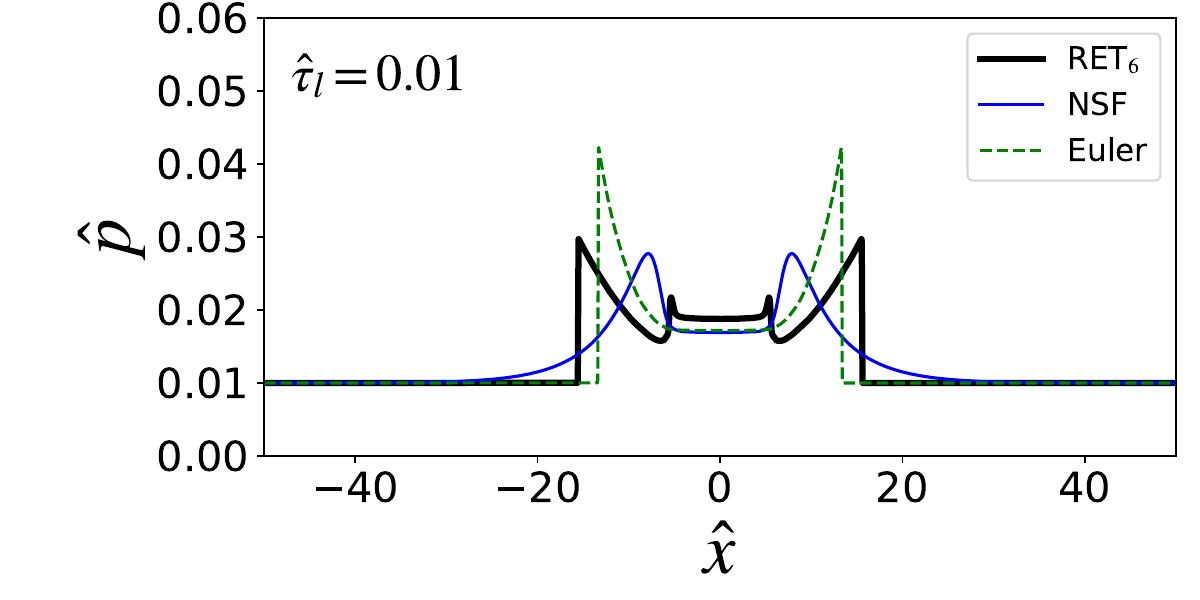}    
    \includegraphics[width=0.33\linewidth]{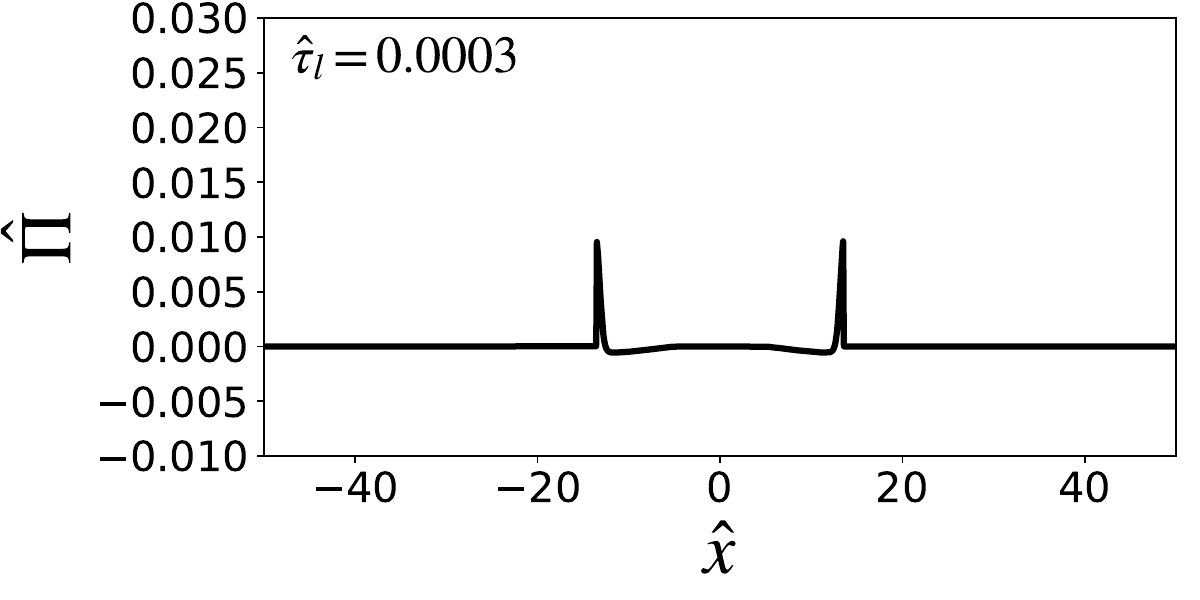}    
    \includegraphics[width=0.33\linewidth]{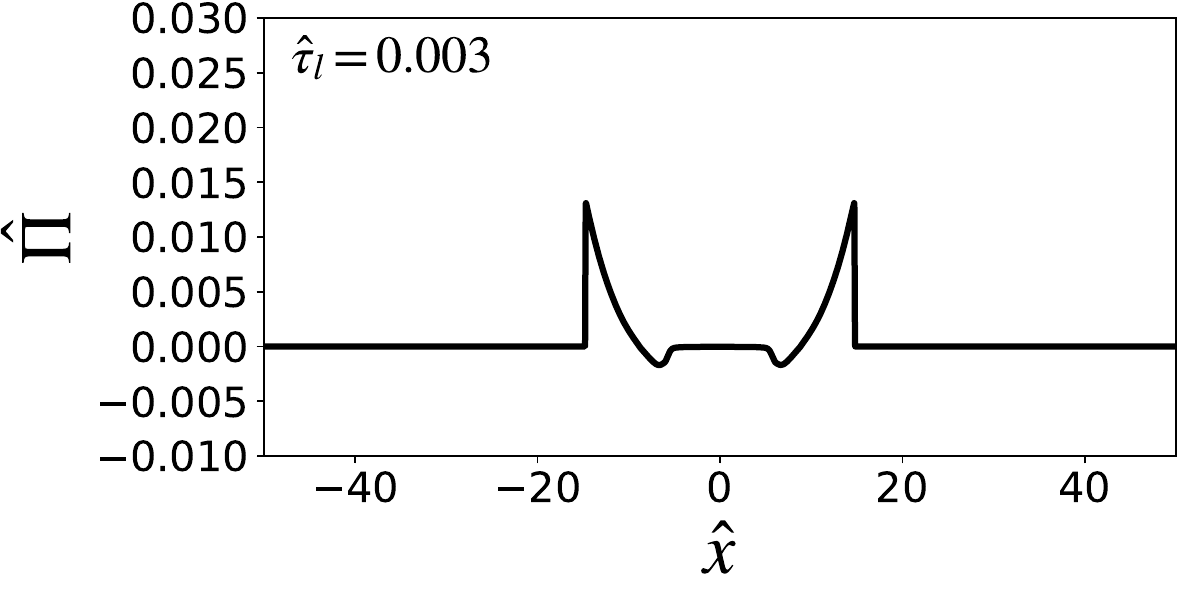}    
    \includegraphics[width=0.33\linewidth]{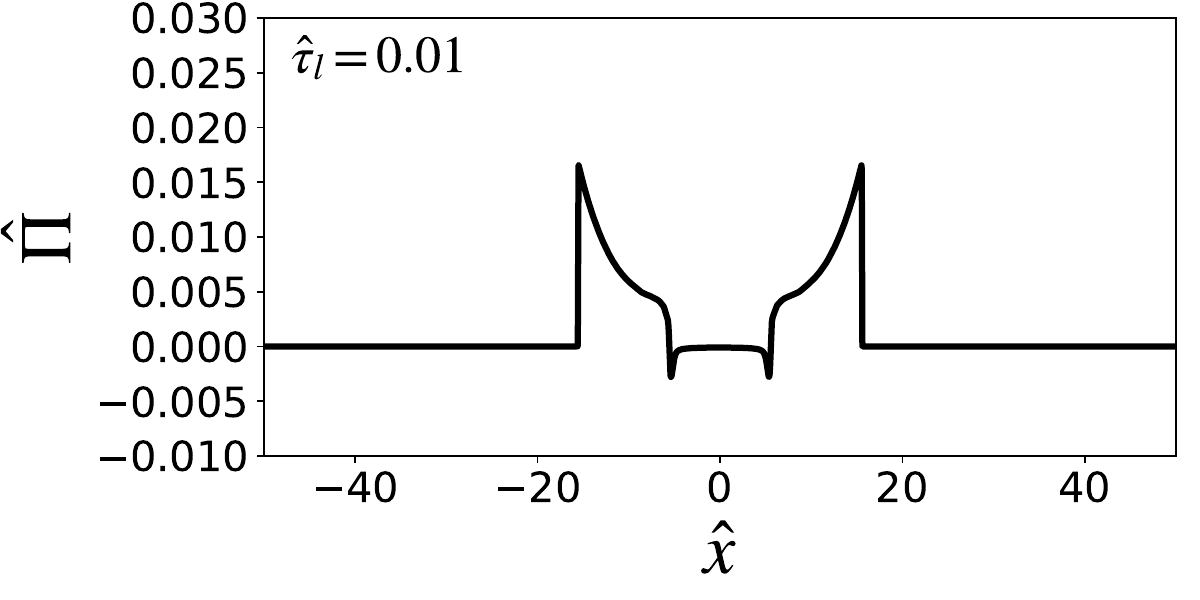}    
    \caption{Profiles of the physical quantities at $\hat{t} = 50$ predicted by the RET$_6$ theory (black thick curves), the NSF theory (blue thin curves), and the system of Euler equations (green dashed curves). 
    The parameters are $\gamma=1.2$, $\hat{L}=0.2$, $\hat{p}_0 = 10^{-2}$, and $n=-1.3$.
    The dimensionless relaxation times are $\hat{\tau}_l=0.0003$ (left column), $\hat{\tau}_l=0.003$ (center column), and $\hat{\tau}_l=0.01$ (right column).}
    \label{fig:gamma1_2_pratio100_th50_n-1_3}
\end{center}
\end{figure*}

Next, we analyze the moderately strong blast waves by adopting $\hat{p}_0 = 10^{-2}$ with a more realistic value of the exponent of the temperature dependence of the bulk viscosity for a carbon dioxide gas $n=-1.3$~\cite{2012_Cramer}. 
We show the profiles of the dimensionless physical quantities at $\hat{t}=50$ for several dimensional relaxation times $\hat{\tau}_I =$ $0.0003$, $0.003$, and $0.01$ in Figure \ref{fig:gamma1_2_pratio100_th50_n-1_3}. 
In addition to the fact that the exponent of $n$ differs from $-1/2$, the present is far from the strong shock limit because the value of $\hat{\rho}_*$ is much less than four. 

Nevertheless, we observe similar behavior discussed in the previous section, although no similarity solution exists in this case. 
All predictions by the RET$_6$ and NSF theories, and the system of the Euler equations, agree with each other when $\hat{\tau}_I$ is small enough. 
With the increase in the value of $\hat{\tau}_I$, the difference between these theories becomes evident, and the RET$_6$ theory predicts significant changes in the (equilibrium) pressure and the dynamic pressure even far from the shock front. 

\section{Summary and concluding remarks}

In this study, we have analyzed the one-dimensional (plane) blast waves based on rational extended thermodynamics with six independent fields (RET$_6$). 
We have derived the similarity solution in the strong shock limit and found the significant deviation from the classical Sedov-von Neumann-Taylor (SvNT) self-similar solution when the dimensionless parameter $\alpha$ becomes large. 
By numerically solving the field equations directly, we have also analyzed the time evolution of the physical quantities and confirmed the validity of the similarity solution derived in the first part. 
It has been shown that the differences between the predictions by the RET$_6$ theory, the corresponding Navier-Stokes-Fourier (NSF) theory, and the system of the Euler equations can become evident even in the moderately strong shock with a generic temperature dependence of the bulk viscosity if the dimensionless relaxation time for the dynamic pressure becomes large. 

In this paper, we have analyzed the simplest plane case with typical examples to demonstrate the critical role of the dynamic pressure on the blast wave. 
A systematic analysis with a variety of parameters, including spherical and cylindrical cases, remains for future subjects. 
The dimensionality effect on the blast waves will soon be reported elsewhere. 

\section*{Declaration of competing interest}
The author has no competing interests.

\section*{Acknowledgment}
This work was supported by Harada Memorial Foundation and Takahashi Industrial and Economic Research Foundation. 
The author thanks Haruto Yamakita and Yui Jibiki for valuable discussions. 

\section*{Data availability}
No data was used for the research described in the article.

\appendix

\section{Derivation of the system for similarity solutions}
\label{sec:derivation}

Let us derive the system for similarity solutions in the same way as the ones adopted in the spherical and cylindrical cases based on the results of the group analysis~\cite{1995_DonatoOliveri,2000_DonatoRuggeri}. 
We require that the system \eqref{ET1D} is invariant with respect to the dilatation group of transformation with $\omega$ ($\neq 0$) being an arbitrary parameter characterizing the dilatation: 
\begin{equation} \label{define}
\begin{split}
&x'=\omega x, \quad t'=\omega^{\phi}t, \quad
\rho'=\omega^{a}\rho, \quad
v'=\omega^{b}v, \quad \\
&p'=\omega^{c}p, \quad
\Pi'=\omega^{d}\Pi, 
\end{split}
\end{equation}
where $\phi$, $a$, $b$, $c$ and $d$ are constants to be determined from the invariant conditions. 
Inserting the quantities defined by \eqref{define} into \eqref{ET1D} and using the invariant condition, we have 
\begin{equation}\label{conditions}
b = 1 - \phi, \qquad c = d = a + 2(1 - \phi). 
\end{equation}
From the invariant condition, the relaxation time also needs to satisfy
\begin{equation*}
\tau (\rho', p') =\omega^{\phi} \tau(\rho, p), 
\end{equation*}
and we obtain the following relationship: 
\begin{equation} \label{zyouken_2}
\begin{split}
&\phi=\frac{2(n-1)-a}{2 n-1}. 
\end{split}
\end{equation}
We notice that the relationships between $\phi$, $a$, $b$, $c$, and $d$ are exactly the same as the ones obtained in spherical and cylindrical cases~\cite{2019_Nagaoka,2021_RdM}. 
Therefore, we adopt the same transformation of the variable as~\cite{2019_Nagaoka,2021_RdM}: 
\begin{equation}\label{ET_henkan_2}
\begin{split}
&\eta=\ln{\xi}=\ln{\frac{x}{t^{\frac{1}{\phi}}}}, \quad \zeta=\frac{1}{\phi}\ln{t},\\
&\rho=\rho_{0}x^{k-a}t^{\frac{a-k}{\phi}} R (\zeta, \eta), \quad 
v=\frac{1}{\phi}\frac{x}{t} \left( V (\zeta, \eta)+1 \right), \\
&p=\frac{\rho_{0}}{\phi^{2}}x^{k-a} t^{\frac{a-k}{\phi}}\frac{x^{2}}{t^{2}} P(\zeta,\eta), \quad
\Pi=\frac{\rho_{0}}{\phi^{2}}x^{k-a} t^{\frac{a-k}{\phi}}\frac{x^{2}}{t^{2}} \Sigma(\zeta,\eta), \\
&\tau=\phi \, t \, \mathcal{T}, 
\end{split}
\end{equation}
where $k$ is another constant parameter to be determined from the invariance condition. 
By inserting \eqref{ET_henkan_2} into \eqref{ET1D}, we have
\begin{align}\label{ET_result}
\begin{split}
&\frac{\partial R}{\partial \zeta} + V \frac{\partial R}{\partial \eta} + R \frac{\partial V}{\partial \eta} 
= -\left\{1+V(k-a+1)\right\}R, \\
&\frac{\partial V}{\partial \zeta} + V \frac{\partial V}{\partial \eta} + \frac{1}{R}\frac{\partial}{\partial \eta}(P+\Sigma) \\
&= - (V+1)(V+1-\phi)-\frac{k-a+2}{R}(P+\Sigma), \\
&\frac{\partial P}{\partial \zeta} + V \frac{\partial P}{\partial \eta} + \left\{ \gamma P + (\gamma-1)\Sigma\right\} \frac{\partial V}{\partial \eta}\\
&= - (V+1)\left[\gamma P + (\gamma-1)\Sigma\right] - \left[2(1-\phi)+V(k-a+2)\right]P, \\
&\frac{\partial \Sigma}{\partial \zeta} + V \frac{\partial \Sigma}{\partial \eta} + \left\{\left(\frac{5}{3}-\gamma\right)P+\left(\frac{8}{3}-\gamma\right)\Sigma\right\}\frac{\partial V}{\partial \eta} \\
&= - \frac{\Sigma}{\mathcal{T}} - (V+1)\left\{\left( \frac{5}{3}-\gamma\right)P+\left( \frac{8}{3}-\gamma\right)\Sigma\right\} \\
&\quad - \left\{ 2(1-\phi) +V(k-a+2)\right\} \Sigma 
\end{split}
\end{align}
with
\begin{equation}\label{TT}
\mathcal{T} = \mathcal{T}_0 x^{2(n-1)-(k-a)}t^{\frac{1}{\phi}(k-a)-2n+1}\frac{P^{n-1}}{R^n}, 
\end{equation}
where $\mathcal{T}_0$ is the following constant: 
\begin{equation*}\label{eq:T_0}
\mathcal{T}_{0}=\frac{\tau_0}{\phi^{2 n - 1}}\left(\frac{p_{0}}{\rho_{0}}\right)^{1-n}. 
\end{equation*}
From the condition that $\mathcal{T}$ is independent of $t$ and $x$, we have the following requirements: 
\begin{equation} \label{zyouken_3}
\begin{split}
k=2(n-1)+a, \qquad 
\phi=\frac{2(n-1)}{2n-1}. 
\end{split}
\end{equation}

We will focus on the similarity solution depending on $x$ and $t$ through a single variable $\xi$, which is the similarity variable in the present analysis.  
In this case, the position of the shock front $X$ can be expressed as 
\begin{equation*}
X = \xi_0 t^{1/\phi}, 
\end{equation*}
where $\xi_0$ is the value of $\xi$ at the shock front. 
By inserting \eqref{ET_henkan_2} into \eqref{E0_def}, we obtain
\begin{equation}\label{E}
\begin{split}
&E_0 = \frac{2 \rho_{0} \, \xi_{0}^{k-a+3}}{\phi^{2} }\int_{0}^{1} \left\{\frac{1}{\gamma-1}P +\frac{1}{2}R (V+1)^{2} \right\} t^{\frac{3-2\phi}{\phi}} \lambda^{k-a+3} d \lambda,
\end{split}
\end{equation}
where $\lambda \equiv \xi/\xi_0$ is the normalized similarity variable such that $\lambda = 1$ holds at the shock front. 
From \eqref{E}, we get
\begin{equation}\label{zyouken_4}
\phi = \frac{3}{2}
\end{equation}
from the condition that $E_0$ should be independent of the time $t$. 
Finally, combining all conditions \eqref{conditions}, \eqref{zyouken_2}, \eqref{zyouken_3}, \eqref{zyouken_4}, we uniquely determine the values of $n$, $a$ and $k$ as 
\begin{equation}\label{conditions2}
\begin{split}
n = -\frac{1}{2}, \quad
&a=0, \quad 
k = - 3.  
\end{split}
\end{equation}
Inserting \eqref{zyouken_4} and \eqref{conditions2} into \eqref{ET_henkan_2}, we determine the dependence of the quantities on $x$ and $t$. 
Furthermore, by dropping the derivatives with respect to $\zeta$, we introduce the dimensionless quantities for obtaining the similarity solution as~\eqref{eq:dimensionless} .


\begin{thebibliography}{00}

%% For numbered reference style

\bibitem{2010_Needham} C. E. Needham, “Blast Waves”, Springer, Berlin Heidelberg (2010). 

\bibitem{1993_Sedov} L.I. Sedov, “Similarity and Dimensional Methods in Mechanics, 10th ed.”, CRC Press, Boca Raton (1993). 

\bibitem{2021_Chakraborti} S. Chakraborti, S. Ganapa, P. L. Krapivsky, and A. Dhar, Phys. Rev. Lett., {\bf 126}, 244503 (2021). 

\bibitem{2021_Ganapa} S. Ganapa, S. Chakraborti, P. L. Krapivsky, and A. Dhar, Phys. Fluids, {\bf 33}, 087113 (2021).

\bibitem{2021_JoyPathakRajesh} J. P. Joy, S. N. Pathak, and R. Rajesh, J. Stat. Phys.,  {\bf 182}, 34 (2021). 

\bibitem{2021_JoyRajesh} J. P. Joy and R. Rajesh, J. Stat. Phys.,  {\bf 184}, 3 (2021). 

\bibitem{2022_KumarRajesh} A. Kumar and R. Rajesh, J. Stat. Phys.,  {\bf 188}, 12 (2022). 

\bibitem{1965_VincentiKruger} W. G. Vincenti and C. H. Kruger Jr., “Introduction to Physical Gas Dynamics”, John Wiley and Sons, New York, London, Sydney, (1965).

\bibitem{1941_BetheTeller}
H. A. Bethe and E. Teller,  
Deviations from Thermal Equilibrium in Shock Waves, reprinted by Engineering Research Institute. University of Michigan (1941). 

\bibitem{1963_deGrootMazur} S. R. de Groot and P.Mazur, “Non-Equilibrium Thermodynamics”, North-Holland, Amsterdam (1963).

\bibitem{1953_GilbargPaolucci}
D. Gilbarg and D.Paolucci,  
%The Structure of Shock Waves in the Continuum Theory of Fluids, 
J. Rat. Mech. Anal. \textbf{2}, 617--642 (1953). 

\bibitem{1998_MullerRuggeri} I. Muller and T. Ruggeri, “Rational Extended Thermodynamics, 2nd ed.”, Springer, New York (1998).

\bibitem{2021_RuggeriSugiyama}	T. Ruggeri and M. Sugiyama, “Classical and Relativistic Rational Extended Thermodynamics of Gases”, Springer, Cham (2021).

\bibitem{2014_PhysRevE} S. Taniguchi, T. Arima, T. Ruggeri, and M. Sugiyama, Phys. Rev. E, {\bf 89}, 013025 (2014). 

\bibitem{2012_CMT} 
T. Arima, S. Taniguchi, T. Ruggeri, and M. Sugiyama, 
Continuum Mech. Thermodyn., \textbf{24}, 271--292 (2012). 

\bibitem{2014_PhysFluids} S. Taniguchi, T. Arima, T. Ruggeri, and M. Sugiyama, Phys. Fluids, {\bf 26}, 016103 (2014).

\bibitem{2016_IJNLM}
S. Taniguchi, T. Arima, T. Ruggeri, and M. Sugiyama, 
Int. J. Non-Linear Mech. {\bf 79}, 66--75 (2016).

\bibitem{2012_PLA}
T. Arima, S. Taniguchi, T. Ruggeri, and M. Sugiyama, 
Phys. Lett. A, {\bf 376}, 2799--2803 (2012). 

\bibitem{2015_IJNLM} 
T. Arima, T. Ruggeri, M. Sugiyama, and S. Taniguchi, 
Int. J. Non-Linear Mech. {\bf 72}, 6--15 (2015). 

\bibitem{2016_Ruggeri} 
T. Ruggeri, 
Bull. Inst. Math. Acad. Sin., {\bf 11}, 1-22 (2016).

\bibitem{2018_BisiRuggeriSpiga}  
M. Bisi, T. Ruggeri and G. Spiga, 
Kinetic and Related Models, {\bf 11}, (1), 71–95 (2018).

\bibitem{2018_KosugeAoki} S. Kosuge and K. Aoki, Phys. Rev. Fluids, {\bf 3}, 023401 (2018). 

\bibitem{2019_KosugeKuoAoki} S. Kosuge, H. W. Kuo, and K. Aoki, 
%Kinetic Model for a Polyatomic Gas with Temperature-Dependent Specific Heats and Its Application to Shock-Wave Structure. 
J. Stat. Phys. {\bf 177}, 209–251 (2019).

\bibitem{1956_IkenberryTruesdell}
E. Ikenberry and C. Truesdell, 
%On the pressure and the flux of energy in a gas according to Maxwell's kinetic theory, 
I. J. Ration. Mech. Anal. {\bf 5}, 1–54 (1956). 

\bibitem{1995_DonatoOliveri} A. Donato and F. Oliveri, Appl. Anal. {\bf 58}, 313-323 (1995). 

\bibitem{2000_DonatoRuggeri} A. Donato and T. Ruggeri, J. Math. Anal. Appl., {\bf 251}, 395-405 (2000). 

\bibitem{2019_Nagaoka} R. Nagaoka, S. Taniguchi, and T. Ruggeri, AIP Conf. Proc., {\bf 2153}, 020014 (2019). 

\bibitem{2021_RdM} S. Taniguchi, Ricerche di Matematica {\bf 70}, 195–206 (2021).

\bibitem{2012_Cramer}
M. S. Cramer, 
%Numerical estimates for the bulk viscosity of ideal gases, 
Phys. Fluids, {\bf 24}, 066102 (2012). 

\bibitem{1998_BoillatRuggeri}
G. Boillat and T. Ruggeri, 
%``On the shock structure problem for hyperbolic system of balance laws and convex entropy,'' 
Cont. Mech. Thermodyn., {\bf 10}, 285--292 (1998). 

\bibitem{2000_LiottaRomanoRusso} S. F. Liotta, V. Romano, and G. Russo, %Central schemes for balance laws of relaxation type, 
SIAM J. Numer. Anal. \textbf{38}, 1337--1356 (2000). 

\bibitem{2006_MentrelliRuggeri}
A. Mentrelli and T. Ruggeri, 
%``Asymptotic behavior of Riemann and Riemann with structure problems for a 2$\times$2 hyperbolic dissipative system,'' 
Suppl. Rend. Circ. Mat. Palermo II, \textbf{78}, 201--225 (2006).

\bibitem{2017_IJNLM}
S. Taniguchi and T. Ruggeri,
%``On the sub-shock formation in extended thermodynamics,'' 
Int. J. Non-Linear Mech. \textbf{99}, 69--78 (2018). 

\bibitem{2017_ConfortoMentrelliRuggeri} F. Conforto, A.  Mentrelli, and T. Ruggeri,  
%``Shock structure and multiple sub-shocks in binary mixtures of Eulerian fluids,''
Ricerche di Matematica, \textbf{66}, 221--231 (2017). 

\bibitem{2022_PoF} T. Ruggeri and S. Taniguchi, Phys. Fluids, {\bf 34}, 066116 (2022). 

\bibitem{2024_CAMC} T. Ruggeri and S. Taniguchi, 
%Effect of Dynamic Pressure on the Shock Structure and Sub-shock Formation in a Mixture of Polyatomic Gases. 
Commun. Appl. Math. Comput., {\bf 6}, 2196–2214 (2024). 

\bibitem{2020_FerzigerPericStreet} J. H. Ferziger, M. Peric, and R. L. Street, “Computational Methods for Fluid Dynamics, 4th ed.”, Springer, Switzerland AG (2020). 

\end{thebibliography}
\end{document}